\documentclass[%
 reprint,
superscriptaddress,
 amsmath,amssymb,
 aps,
]{revtex4-1}

\usepackage{graphicx}
\usepackage{dcolumn}

\usepackage{subcaption}
\usepackage{tabularx}
\usepackage{multirow}
\usepackage[dvipsnames]{xcolor}
\usepackage[export]{adjustbox}
\usepackage{cleveref}
\usepackage{ragged2e}
\makeatletter
\renewcommand{\@makecaption}[2]{%
  \vskip\abovecaptionskip
  \footnotesize\noindent\justifying \textbf{#1}: #2\par
  \vskip\belowcaptionskip
}
\makeatother

\makeatletter
\newcommand{\FigPanels}[2]{%
  Fig.~\ref{#1}#2%
}
\makeatother

\usepackage{dsfont}
\usepackage{bm}

\begin{document}

\preprint{APS/123-QED}

\title{Conformal Elastodynamics in $2$D Dilational Metamaterials}

\author{Neel Singh}
\thanks{These authors contributed equally to this work.}
\affiliation{%
School of Physics, Georgia Institute of Technology, Atlanta, Georgia, 30332, USA
}%

\author{Audrey A. Watkins}%
\thanks{These authors contributed equally to this work.}
\affiliation{%
 School of Engineering and Applied Science, Harvard University, Cambridge, MA 02138
}%
\author{Giovanni Bordiga}
\affiliation{%
 School of Engineering and Applied Science, Harvard University, Cambridge, MA 02138
}%

\author{Vincent Tournat}
\affiliation{%
 Laboratoire d’Acoustique de l’Université du Mans (LAUM), UMR 6613, Institut d’Acoustique - Graduate School (IA-GS), CNRS, Le Mans Université, France
}%
\affiliation{%
 School of Engineering and Applied Science, Harvard University, Cambridge, MA 02138
}%

\author{Katia Bertoldi}
\affiliation{%
 School of Engineering and Applied Science, Harvard University, Cambridge, MA 02138
}%
\author{Zeb Rocklin}
\thanks{Corresponding author. Email: zebrocklin@gatech.edu}
\affiliation{%
School of Physics, Georgia Institute of Technology, Atlanta, Georgia, 30332, USA
}

\date{\today}

\begin{abstract}
Flexible mechanical structures can undergo large deformations under small loads, enabling large, complex, and nonlinear wave responses under finite-frequency driving. Here, we study a dynamically driven canonical flexible mechanical metamaterial composed of rigid squares connected at their corners by flexible hinges. This metamaterial supports a uniform dilational mechanism and, in the limit of ideal joints, exhibits a Poisson ratio of $-1$. The presence of this dilational mode of deformation gives rise to a conformal symmetry, in which the dynamics are approximately invariant under a wide class of physical transformations---conformal maps. We find that the low-frequency response of the system is dominated by conformal deformations consisting of spatially varying rotations and dilations concentrated at the boundary. Even at high frequencies, each conformal map implies a conserved spatially complex momentum. We explore how experimental parameters such as material stiffnesses and the geometry and number of unit cells allow experimental conformal momenta to approach this conservation, varying slowly compared to the non-conformal momenta of same order. These results constitute a new framework opening fundamental avenues for the study of conformal wave phenomena in dilational metamaterials as well as potential strategies for controlling nonlinear waves and vibrations.
\end{abstract}

\maketitle

\section{\label{sec1}Introduction}

Flexible structures capable of large, reversible changes in form are ubiquitous across natural and engineered systems.
Examples range from snapping of the Venus flytrap \citep{Forterre2005} and shape-morphing cellular membranes \citep{Lipowsky1991}, to engineered deployable structures spanning everyday mechanisms like umbrellas, medical devices such as vascular stents \citep{Pan2021}, space-borne deployable architectures \citep{Gardner2006,packaging_miura_1985}, and other programmable structures \citep{Melancon2021,crushing_wierzbicki_1983, Buckliball2013}.
Mechanical metamaterials exploit flexibility engineered into their microstructure to access elastic behaviour far beyond that of conventional solids \citep{Bertoldi2017}, including zero or negative effective material properties \citep{lakes87, Nicolaou2012,BERTOLDI2008, Kadic2012,mechanical_Yu_2017,Berger2017,Shim_auxetic_2013,Coulais2017}, as well as nonlinear \citep{packaging_miura_1985,Cho2014,Coulais2015,Deng2017,Deng2020,Deng2021}, programmable \citep{Mullin2007,Silverberg2014,Florijn2014,Coulais2016,Paulose2023}, and topological responses \citep{zebtopomech,kanelubetopomech,VitelliTopoSoliton2014,Huber2016,Saremi2020}.
Furthermore, mechanical metamaterials operating in the dynamic regime can be deliberately engineered to control nonlinear waves \citep{deng2021nonlinear}. Such wave control phenomena include unidirectionally propagating solitons \citep{raney2016stable}, mechanical cloaking \citep{bordiga2025nonlinear}, and energy focusing \citep{Carrara2013,Bordiga2024}. By assembling rigid elements connected by soft joints, these systems provide an unusual degree of control over macroscopic mechanical response, enabling systematic design and characterisation of elasticity from geometry alone. Yet despite this architectural tunability, it remains difficult to formulate universal principles governing their strongly nonlinear and dynamic behaviour.

\begin{figure*}[t]
\centering
\includegraphics[width=\textwidth]{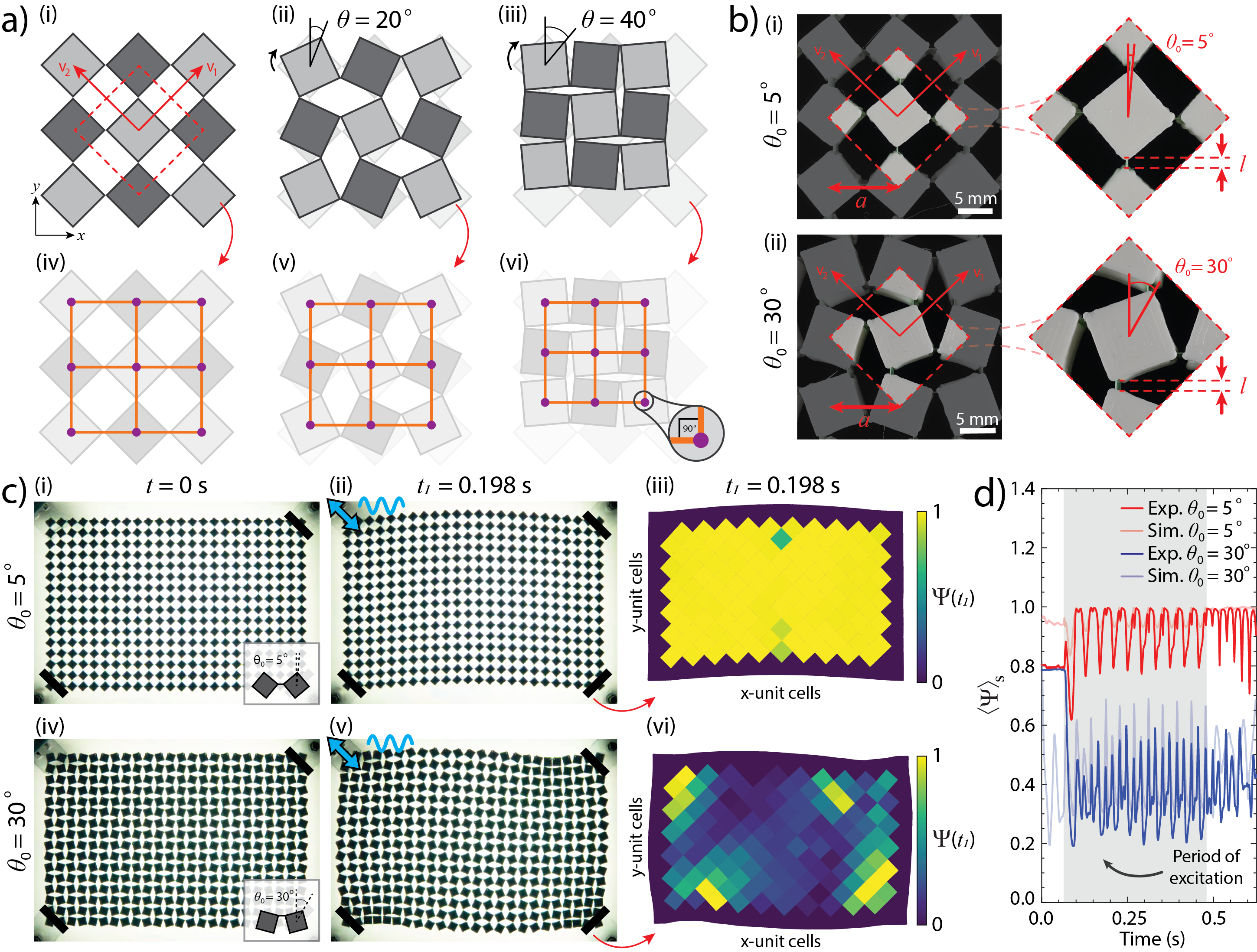}
\caption{\textbf{RS lattice description}:
\textbf{(a)}(i-iii) The rotating square lattice undergoes a mechanism in which rigid squares counter-rotate through angles $\pm \theta$. (iv-vi) This motion leads to a dilation without shear of the system, with the grid changing area while maintaining shape. 
\textbf{(b)} Samples are fabricated with block spacing distance $a$, connected via hinges of length $l$, and rotated by initial angles $\theta_0$ in alternating directions relative to the fully expanded state.
\textbf{(c)} (i,iv) The two samples are suspended by their corners and (ii,v) driven at one corner by an oscillating in-plane shaker. (iii, vi) In the first sample the shear to total strain ratio is significant across the sample. In contrast, the second sample has very little shearing, characteristic of conformal deformations.
\textbf{(d)} The time-evolution of the spatially averaged shear to total strain ratio in response to harmonic excitation of the samples in experiment and simulation in response to the experimental drive. }
\label{fig:figure1}
\end{figure*}

$2$D dilational metamaterials form a subclass characterized by a maximally auxetic soft dilational mode \citep{auxeticrs, Bertoldi2010, Shan2015}. A canonical example is the Rotating-Square (RS) design shown in \FigPanels{fig:figure1}{a}, whose low-energy deformations under quasi-static conditions correspond to non-uniform conformal maps \citep{michaelconformal}, mathematical functions that encode locally dilational shape changes (local expansion or contraction) \citep{england2003complex}.
Because these deformations can occur continuously with negligible energy cost, there is an inherent ambiguity in defining a `true' reference state. This reflects an underlying conformal symmetry, a highly unusual feature for elastic solids \citep{RIVA2005,Baggioli2020a,Baggioli2020b}.
Although recent studies of the static response of dilational metamaterials have noted aspects of this symmetry \citep{michaelconformal, czajkowski2022duality, Sun2012, ian_paul_conformal}, its deeper implications for dynamic elastic behaviour remain largely unexplored.

In this work, we explore this connection between flexibility and symmetry in the context of the dilational RS metamaterial, to build a framework for understanding and controlling the dynamic response. 
We show how the long-wavelength dynamics are approximately described by a reduced-order conformal elasticity theory, with perturbative deviations accounting for the small energy cost of dilation and finite-size effects.

In the ideal limit where perturbations are negligible, the conformal dynamics remain invariant under infinitesimal conformal transformations of the reference state.
Thus, the ability to deform conformally establishes a novel conformal symmetry. In analogy to the usual translation and rotational symmetries of solids \citep{landau1986theory}, this new symmetry implies that the local size is no more constrained than position or orientation in dilational systems.
We apply Noether’s theorem \citep{Noether1918,spontaneousbreak}, a fundamental result in physics, to derive new conservation laws for generalised momenta associated with conformal maps beyond linear and angular momenta.

Through theory, numerical simulations and experiments on fabricated RS metamaterials, we predict and verify the implications of conformal symmetry for realistic dilational metamaterials.
Although this symmetry is weakly broken in real systems by perturbative terms, their dynamics remain strongly influenced by proximity to the ideal conformal model.
As a result, the low-frequency response is dominated by near-conformal boundary modes that reduce to zero-frequency modes in the ideal limit.
The novel conserved conformal momenta derived from Noether’s theorem remain approximately conserved in real dilational systems, even at high frequencies.
Finally, we identify the key experimental parameters controlling the magnitude of these effects, namely the ratio of effective bulk to shear modulus and the metamaterial size. 

\section{\label{sec2} System}

\FigPanels{fig:figure1}{a} depicts the rotating square (RS) metamaterial, which consists of a network of rigid squares connected at their corners by rotational joints. In the limit of freely bending corner joints (ideal hinges), the structure can deform via a zero-energy mechanism where adjacent squares counter-rotate by the same angle $\theta$ in alternating directions, resulting in uniform expansion or contraction of the lattice (\FigPanels{fig:figure1}{a(i-iii)}). As evident from tracking an overlaying grid defined by the square centroids, shown in orange in \FigPanels{fig:figure1}{a(iv-vi)}, this mechanism follows a coarse uniform dilation that only affects the area of the grid but preserves its local shape. Finally, we note that the primitive unit cell of the RS metamaterial is a square centred at the centroid of a rigid block, with its edges aligned with the lattice vectors $\mathbf{v}_1 = a ( \hat{\imath} + \hat{\jmath})$ and $\mathbf{v}_2 = a (-\hat{\imath} + \hat{\jmath})$, where $a$  is the distance between adjacent block centroids and $\hat{\imath}$ and $\hat{\jmath}$ are unit vectors aligned with the $x$ and $y$ directions, respectively (\FigPanels{fig:figure1}{a}).

We fabricate RS metamaterials  using 3D-printed polylactic acid (PLA) units connected by thin polyester plastic shims, which act as flexible hinges. During deformation, these finite-length shims experience not only bending but also stretching and shear, which compete with the idealised pure dilation mode of the mechanism. To ensure that stretching and shear do not dominate the response, we design the shims to be sufficiently thin and short, with a length of $l = 0.5$ mm and a thickness of $h = 76.2~\mu$m.  We consider two samples, both consisting of a  $24 \times 16$ array of units with a centre-to-centre distance of $a = 10$ mm, but zero-energy states defined by an initial angle of either $\theta_0 = 5^\circ$ or $\theta_0 = 30^\circ$ (\FigPanels{fig:figure1}{b}). 

In our tests, we suspend each sample in the air by clamping three corner blocks at three of the four corners of the structure (\FigPanels{fig:figure1}{c}) and dynamically excite by applying a harmonic excitation with amplitude $A$ and frequency $f_{\mathrm{d}}$ via a low-frequency shaker attached to the fourth corner at a $315^\circ$ angle. The response is recorded using a high-speed camera, and a tracking algorithm based on digital image correlation is employed to reconstruct the displacement field of all units.  We find that under a harmonic excitation signal with frequency $f_{\mathrm{d}} = 24$ Hz and amplitude $A = 4$ mm, the two samples exhibit visibly different dynamic responses. To illustrate these differences, in \FigPanels{fig:figure1}{c} we present snapshots of both structures at $t= 0.198$~s. In the case of the $\theta_0 = 5^\circ$ sample, minimal dilation is observed while the sample is excited. In contrast, the $\theta_0 = 30^\circ$ sample exhibits nearly pure dilational behaviour with minimal shearing.

To evaluate the dilational and shear responses of the two samples, we begin by computing the position of each unit cell as the weighted average of the displacements of its constituent blocks. This is then used to obtain the deformation gradient $\mathbf{F}$ for each unit cell (see Supplementary Information, Section  III). Finally, we calculate the nonlinear dilational and shear strain magnitudes for each unit cell as
\begin{equation}
    d = \sqrt{\text{Det}[\textbf{C}]} -1 ,
    \label{eqn:dilation}
\end{equation}
\begin{equation}
    s^2 = \text{Tr[\textbf{C}]} - 2\sqrt{\text{Det[\textbf{C}]}} ,
    \label{eqn:shear}
\end{equation}
where $\textbf{C}=\textbf{F}^T\textbf{F}$ is the right Cauchy-Green deformation tensor (see Supplementary Information, Section  III) \cite{michaelconformal}. In \FigPanels{fig:figure1}{c(iii)} we report  the spatial distribution of the  ratio of shear strain to total strain 
\begin{equation}
    \Psi(t) = \sqrt{\frac{s^2(t)}{s^2(t)+d^2(t)}},
    \label{eqn:psi}
\end{equation}
 at $t_1 = 0.198$ s, revealing strikingly different spatial patterns. The sample with $\theta_0 = 5^\circ$ exhibits $\Psi$ values that approach $1$, indicating high shearing and minimal dilating of unit cells throughout most of the bulk. In contrast, the sample with $\theta_0 = 30^\circ$ displays much lower values of $\Psi$, indicating minimal shearing with dominant dilational strains across the bulk, with only slight shearing occurring near the clamped corners (\FigPanels{fig:figure1}{c(vi)}).

\begin{figure*}[t]
\centering
    \begin{subfigure}[t]{\linewidth}
        \includegraphics[width=\textwidth]{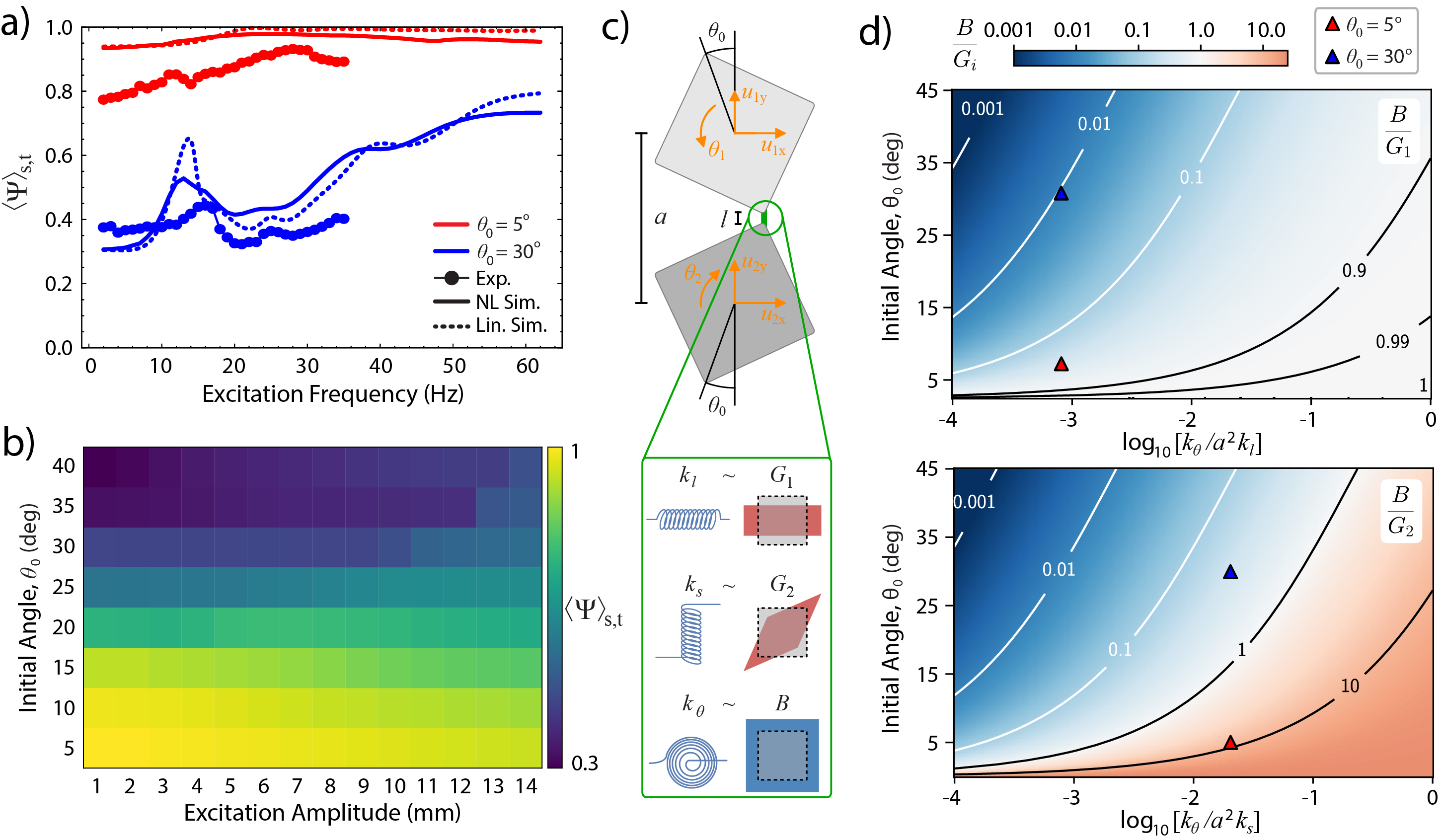}
        \label{fig:linearmodel}
    \end{subfigure}
\caption{\textbf{(a)} Evolution of  $\langle \Psi \rangle_{\text{s,t}}$ as excitation frequency is varied in experiments, nonlinear simulations, and linear simulations for $\theta_0 = 5^{\circ}$ and $\theta_0=30^{\circ}$. \textbf{(b)} Evolution of $\langle \Psi \rangle_{\text{s,t}}$ as $\theta_0$ and excitation amplitude is varied. \textbf{(c)} Schematic of reference configuration of two adjacent blocks with spacing $a$, rotated by $\theta_0$ in alternating directions relative to fully-expanded state, with $3$ degrees-of-freedom (orange): $2$ displacements and an in-plane rotation measured in opposite conventions for squares $1$ (light grey) and $2$ (dark grey); their corners connected through soft ligaments (green) of length $l$, modelled as $3$ springs resisting stretching, shearing and bending (green box). The spring stiffnesses relate to effective elastic moduli in the continuum limit: two shear moduli $G_1$ and $G_2$, and bulk modulus $B$. \textbf{(d)} The ratios of bulk-to-shear moduli, $B/G_{1/2}$, are plotted against design parameters, revealing how dilation is energetically favourable to shears, $B < G_{1/2}$, for large initial angle $\theta_0$, and easy-to-bend ligaments, i.e. small non-dimensionalised stiffness ratios $k_{\theta}/a^2k_{s/l} \ll 1$.}
\label{fig:figure2}
\end{figure*}

Next, in \FigPanels{fig:figure1}{d}, we show the spatial average of $\Psi$, denoted as $\langle \Psi \rangle_s$, for both samples as a function of time. We find that the shear-to-total strain ratio of the sample with $\theta_0 = 5^\circ$ remains nearly twice that of the sample with $\theta_0 = 30^\circ$ during the tests. These clear quantitative differences in dynamic response reveal that the $\theta_0 = 30^\circ$ sample deforms primarily through dilation, whereas the $\theta_0 = 5^\circ$ sample exhibits shear-dominated behaviour. As shown in \FigPanels{fig:figure2}{a}, these distinctions persist across the entire tested frequency range up to 35 Hz, underscoring the robustness of the observed behaviour.

To elucidate the qualitative differences between the two metamaterial configurations, we begin by modelling their low-energy dynamics using discrete rigid squares with three degrees of freedom that are free to translate and rotate in-plane  (see \FigPanels{fig:figure2}{c}). These squares are connected at their vertices by three independent linear springs that resist axial, shear, and bending deformations, with effective stiffnesses of $k_l = 20.59$ N/mm, $k_s = 0.83$ N/mm, and $k_\theta = 1.68$ N·mm, respectively. As shown in \FigPanels{fig:figure1}{d} and \FigPanels{fig:figure2}{a}, the numerical predictions of $\langle \Psi \rangle_s$ are in close agreement with the experimental observations, confirming the validity of this model.

While the results in Fig.~\ref{fig:figure1} provide insight into the evolution of $\Psi$ under a fixed dynamic input with amplitude $A = 4$ mm for samples with $\theta_0 = 5^{\circ}$ and $\theta_0 = 30^{\circ}$, we next use the model to systematically explore how $\Psi$ evolves as a function of both $A\in[1,14]$ mm and $\theta_0\in[5^{\circ},\, 40^{\circ}]$. \FigPanels{fig:figure2}{b} highlights the strong dependence on the initial angle $\theta_0$ of the space- and time-averaged shear 
to total strain ratio, $\langle \Psi \rangle_{\text{s,t}}$, where $\langle \cdot \rangle_{\text{t}}$ is averaged across 10 periods of excitation. In contrast, $\langle \Psi \rangle_{\text{s,t}}$ exhibits only a weak dependence on the excitation amplitude, remaining nearly constant across the considered amplitude range. This observation indicates that the dynamics of the metamaterial can be analysed in the small-amplitude limit. As shown in \FigPanels{fig:figure2}{a}, the linearized discrete model (described in Supplementary Information, Section IV S3) suffices to capture the behaviour observed in the nonlinear experiments.


\section{\label{sec3} Continuum Model}

In order to better understand the origin of dilation-dominated low-frequency vibrations of \FigPanels{fig:figure2}{a} for the sample with $\theta_0=30^\circ$, we develop an effective continuum linear elastic theory by coarse-graining the linearized discrete spring-mass model, and analyse the geometric and material conditions under which dilations dominate the low-frequency response. Such homogenisation is a standard approach for characterising the response of complex mechanical lattices \citep{Milton_2002,GONELLA2008459,dellIsola2016,Lestringant2023,AUDOLY2026105956}.

As shown in \FigPanels{fig:figure2}{c}, each unit cell is comprised of two blocks with six degrees of freedom per cell. In the continuum description, these are promoted to six continuum fields defined over space $\mathbf{r}$ and time $t$, interpolating the block displacements and rotations across all cells. At low energy, we a priori expect two acoustic dispersion bands that describe the dynamics in terms of the slowly-varying mean displacement field $\mathbf{u}(\mathbf{r},t)$, extending from zero-frequency uniform translations $\mathbf{u} = \mathrm{const}$.
In the long-wavelength limit for these acoustic bands, the other four degrees of freedom relax in relation to $\mathbf{u}$, as found by perturbing the system around the uniform translation modes and minimising the deformation energy (see  Supplementary Information, Section V).
Any deviations from this relaxed manifold are gapped away at high frequency, and therefore not excited in the low-frequency regime.

One such deviation is the field of relative displacements between neighbouring blocks, which induces shearing and stretching in the hinges.
The corresponding energy cost scales with the amplitude of this excitation itself, rather than with its spatial gradients. Consequently, such excitations remain at finite frequency even in the long-wavelength limit, effectively removing two degrees of freedom, the relative displacements along $x$ and $y$, from the low-frequency sector.

Excitations in the mean displacement field $\mathbf{u}$ deform hinges and generate forces only at the order of its gradients, allowing these two degrees of freedom in the low-frequency sector. The gradient components of $\mathbf{u}=(u_x,u_y)$ are defined as:
\begin{align} \label{eqn:linearstrain&rotn}
    \begin{aligned}
    d & := \partial_x u_x + \partial_y u_y, & s_1 & := \partial_x u_x - \partial_y u_y, \\
    r & := \partial_x u_y - \partial_y u_x, & s_2 & := \partial_x u_y + \partial_y u_x.
    \end{aligned}
\end{align}
where $r$ is the coarse-grained rotation and $d,s_1,s_2$ are the $2$D coarse-grained strains, linearised versions of the nonlinear definitions in \cref{eqn:dilation,eqn:shear} with $s^2 = s_1^2 + s_2^2$. Note that the rotational component $r$ does not deform the hinges provided the blocks rotate consistently with it, a consequence of the rotational symmetry inherent to any mechanical system. This links the mean in-phase block rotation between neighbouring squares to the macroscopic rotation field $r$. Deviations from this correspondence shear the hinges, and are therefore gapped away at high frequency, removing a rotational degree of freedom from our analysis.

Similarly, the counter-rotations of neighbouring blocks that cause bending in the common hinge couple with the local dilation $d$, with deviations gapped away at high frequency. In the ideal limit of freely-bending hinges of negligible size, the geometry gives a linear relationship between this counter-rotation field $\theta_d$ and the coarse-grained dilation field $d$ with a proportionality factor dictated by the reference angle $\theta_0$, such that a fully expanded structure ($\theta_0=0^\circ$) does not dilate to linear order in counter-rotation. For the real hinges, there are small corrections associated with their finite size $l$ and the finite stiffness to their bending $k_{\theta}$, which induce the hinges to somewhat stretch/compress under macroscopic dilation in combination with bending, leading to the following expression for dilation:
\begin{align}
    d(\mathbf{r}) & = - 2\theta_d(\mathbf{r}) \tan{\theta_0} + \mathcal{O}(l/a) + \mathcal{O}(k_{\theta}/ a^2 k_l). \label{eqn:idealmechanism}
\end{align}
The detailed derivation can be found in Supplementary Information, Section V.
This relation highlights that lattices with larger $\theta_0$ can accommodate greater dilation through hinge bending than those with smaller $\theta_0$. Accordingly, in our experimental samples, the design with $\theta_0=30^{\circ}$  is expected to easily accommodate macroscopic dilation via low-energy microscopic hinge-bending compared to the $\theta_0=5^{\circ}$ design.

With the six degrees of freedom reduced to two on the acoustic bands, the deformation energy density is found to be quadratic and decoupled in the linear coarse-grained strains, $d$, $s_1$ and $s_2$, ignoring higher-order gradients; see Supplementary Information, Section V S2, for details on the derivation. Thus, the low-energy dynamics of the RS metamaterial are described by an effective Lagrangian density that represents a linear $2$D orthotropic elasticity,
\begin{align}
    \mathcal{L} & = \frac{\rho}{2} |\dot{\mathbf{u}}|^2 - \left ( \frac{B}{2} d^2 + \frac{G_1}{2} s_1^2 + \frac{G_2}{2} s_2^2 \right )+ \mathcal{O}\left (\frac{G_1}{N}\right) , \label{eqn:continuumLagrange}
\end{align}
where $\rho = m/a^2$ is the coarse-grained mass density for block mass $m$, $B$ is the bulk modulus, and $G_1,\, G_2$ are the two shear moduli, which are related to the microscopic hinge stiffnesses:
\begin{align}
    B & \approx \frac{2}{\tan^2\theta_0} \frac{k_{\theta}}{a^2} +  \mathcal{O} \left (\frac{l}{a} \frac{k_{\theta}}{a^2} \right ) + \mathcal{O}\left (\frac{k_{\theta}}{k_l a^2} \frac{k_{\theta}}{a^2} \right ), \nonumber \\
    G_1 & = \frac{k_l}{2}, \qquad \qquad \qquad
    G_2 = \frac{k_s}{2}. \label{eqn:elasticmoduli}
\end{align}

The shear-moduli expressions reflect how linear shear strains map onto hinge stretching and shearing deformations (see \FigPanels{fig:figure2}{c}), as dictated by geometry. Generally, the bulk modulus depends on both bending and stretching stiffness. However, for ideal zero-size unstretchable hinges, dilation occurs entirely via bending as described in \cref{eqn:idealmechanism}, and the bulk modulus only depends on the bending stiffness $k_{\theta}$, as approximated in \cref{eqn:elasticmoduli}.
While these elastic moduli are comparable in conventional elastic materials, we find that the RS metamaterial exhibits a bulk modulus much lower than shear moduli for large initial angles $\theta_0$, as shown in \FigPanels{fig:figure2}{d}.
Therefore, dilations are energetically favoured over shear strains for large-angle designs, whereas small-angle designs favour shear. This qualitatively explains the dilation-dominant response of the $30^{\circ}$ sample compared to the $5^{\circ}$ sample.

\begin{figure*}[t]
\centering
    \begin{subfigure}[t]{\linewidth}
        \includegraphics[width=\textwidth]{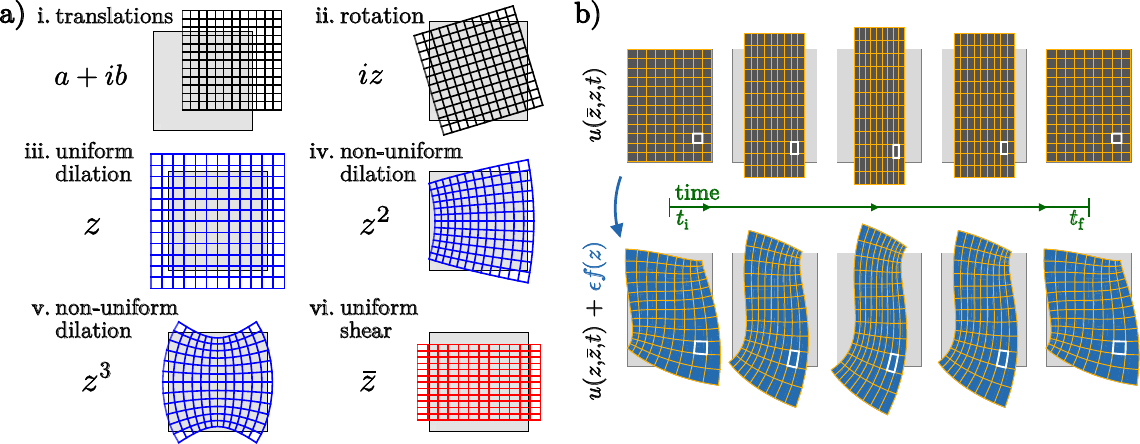}
    \end{subfigure}
\caption{\textbf{(a)} Mechanical structures in general have dynamics that are invariant under translations (i) and rotations (ii), whereas metamaterials that dilate freely can also undergo deformation through conformal maps of uniform dilations (iii) and even spatially varying dilations (iv,v). In contrast, other (non-conformal) deformations such as uniform shear (vi) generate stress and cost energy.
\textbf{(b)} In the top row, a mechanical system undergoes a uniform shearing motion. In the bottom row, the system undergoes the same trajectory with an additional static conformal displacement. For systems that dilate freely, both trajectories have the same Lagrangian and action, making static conformal deformations symmetries of such systems. Because the original deformation includes shear, the highlighted region changes shape over time, however the addition of the conformal map only translates, rotates and dilates the region, without introducing additional shear strains, such that forces and energies are unchanged.
}
\label{fig:figure3}
\end{figure*}

\begin{figure*}[t]
\centering
    \begin{subfigure}[t]{\linewidth}
        \includegraphics[width=\textwidth]{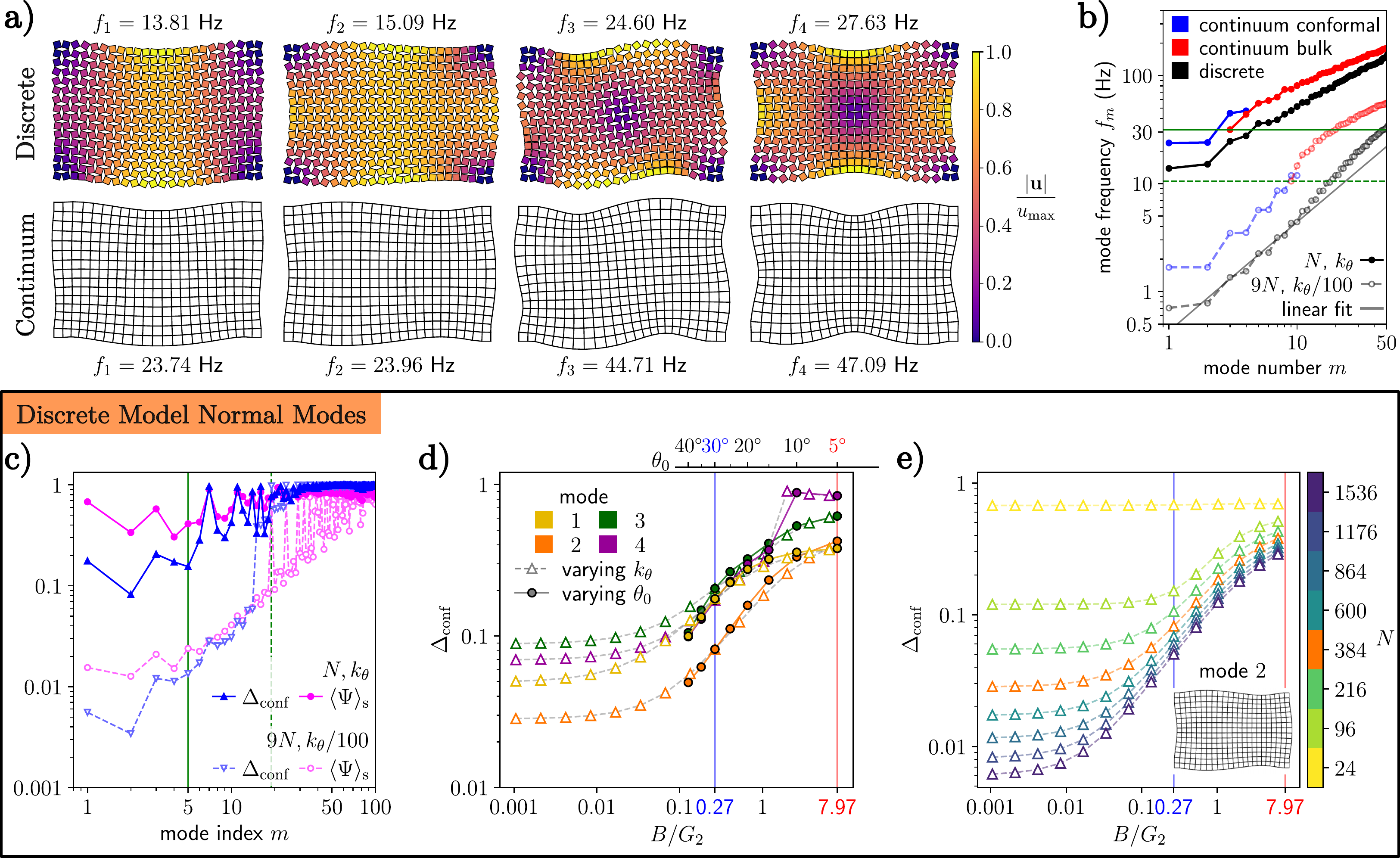}
        \label{fig4}
    \end{subfigure}
\caption{
First $4$ predicted conformal-mode profiles for fabricated $\theta_0=30^{\circ}$ sample with $N=24\times 16$ blocks and bending stiffness $k_{\theta}=$ are shown in \textbf{(a)} with agreement between discrete and continuum theory.
Corresponding dispersion plot in \textbf{(b)} shows agreement between discrete-model predictions (black) and continuum predictions of conformal modes (blue) below non-conformal bulk modes (red), with more conformal modes predicted at even lower frequencies for an improved lattice design (dashed), with more blocks and softer hinges. The frequency of the first continuum bulk mode sets the ceiling (green) for conformal behaviour, $f_{\mathrm{ceil}}$. We fit conformal polynomials to these modes and plot the normalised fitting error $\Delta_{\mathrm{conf}}[m]$ in \textbf{(c)} (blue), verifying modes below this ceiling (green line) are close-to-conformal, with the improved lattice design exhibiting even lower fitting errors. Concurrently, the lower-frequency modes also have smaller fractional shear $\langle \Psi \rangle_{\mathrm{s}}$ than the higher-frequency modes, with the improved lattice design showing negligible shearing.
\textbf{(d)} The minimised conformal fitting error for first $4$ modes of the $N=24\times 16$ lattice, predicted by discrete model, decreases with decreasing $B/G_2$ ratio, which is controlled by two parameters, $\theta_0$ and $k_{\theta}$, plotted separately. The two experimental cases with $\theta_0 = 5^{\circ}$ and $30^{\circ}$ are highlighted in red and blue, respectively. As $B/G_2$ gets smaller, fitting error saturates due to lattice effects becoming more prominent, as verified in \textbf{(e)} that tracks the fitting-error-vs-$B/G_2$ trend in mode $2$ for different lattice sizes $N$ of same aspect ratio $3:2$, showing how the deformation saturates to a higher conformal nature (lower fitting-error) for larger $N$.}
\label{fig:figure4}
\end{figure*}

\begin{figure*}
\centering
    \begin{subfigure}[t]{\linewidth}
        \includegraphics[width=\textwidth]{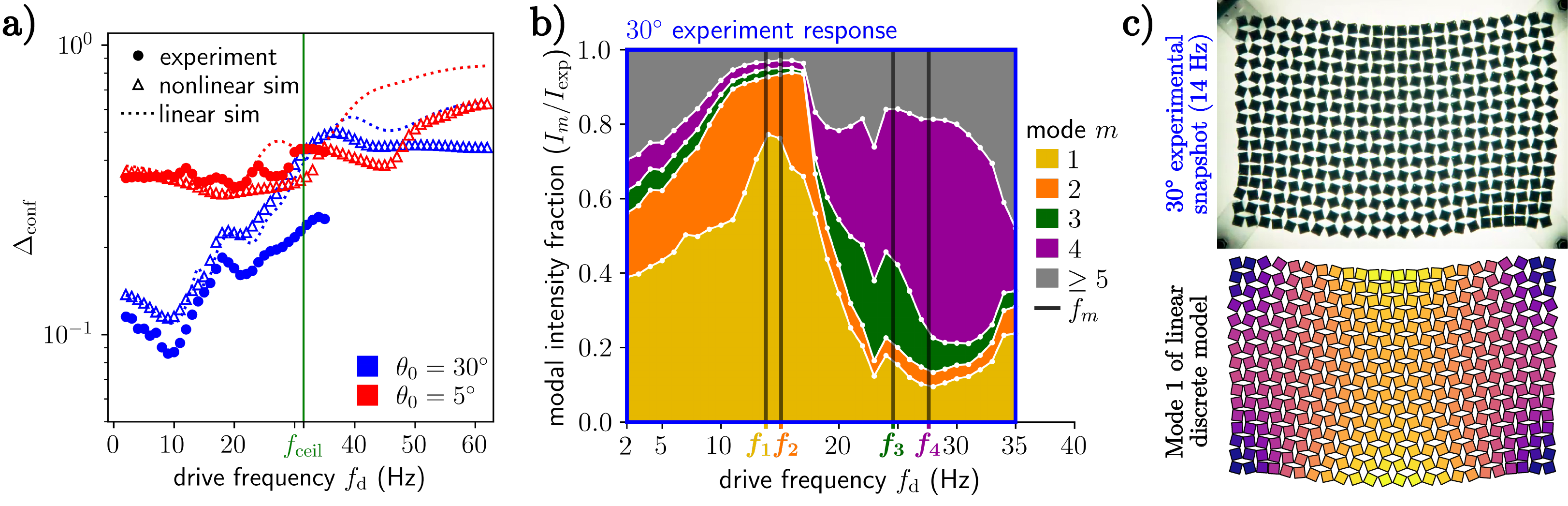}
    \end{subfigure}
\caption{\textbf{(a)} Tracked displacement response for corner-suspended experiments shows significantly smaller non-conformal deviations in $30^{\circ}$ (blue) sample as compared to $5^{\circ}$ (red) sample at low frequencies, steadily increasing as drive frequency approaches $f_{\mathrm{ceil}}$ exciting more non-conformal bulk waves, validating the predictions by linear and nonlinear discrete simulations.
\textbf{(b)} shows the decomposition of the $30^{\circ}$-experiment response with the normalised modal intensities along first $4$ modes and the remaining deformation intensity along other modes stacked along $y$-axis, verifying that the response is mostly exciting the predicted low-frequency conformal modes.
\textbf{(c)} shows an experimental snapshot visually similar to the mode-$1$ profile found numerically for linear discrete model, due to driving close to resonance ($f_{\mathrm{d}}=14 \mathrm{Hz} \approx f_1$).
}
\label{fig:figure5}
\end{figure*}

\section{\label{sec4}Conformal Mechanics}

While the low bulk modulus seems to permit dilational modes at low frequency, the conventional continuum theory does not readily describe how the non-uniform dilations of \FigPanels{fig:figure1}{c,d} are possible. To address this, we shift to a complex field theory similar to those used in recent works on two-dimensional dilational systems~\citep{michaelconformal, Sun2012, ian_paul_conformal}, and work demonstrating the absence of conformal symmetry in conventional systems~\citep{RIVA2005}.
The complex formulation maps real two-vectors, such as positions $\mathbf{r}=(x,y)$ or displacements $\mathbf{u}$, to complex scalars such as $z \equiv x+i y$ and their complex conjugates, denoted $\bar{z}$.
The gradients of the displacement field---dilation $d$, rotation $r$ and shears $s_1, s_2$ [\cref{eqn:linearstrain&rotn}]---emerge naturally using standard complex derivatives described in Supplementary Information, Section VI S1:
\begin{align}
    \begin{aligned}
    2\partial_z u & = \partial_x u_x + \partial_y u_y + i(\partial_x u_y - \partial_y u_x) = d + ir , \\
    2\partial_{\bar{z}} u & = \partial_x u_x - \partial_y u_y + i(\partial_x u_y + \partial_y u_x) = s_1 + is_2.
    \end{aligned} \label{eqn:complexstrains}
\end{align}
Thus, the holomorphic derivative $\partial_z u$ encodes the shape-independent gradients, dilation and rotation, while the anti-holomorphic derivative $\partial_{\bar{z}} u$ encodes the shape-change gradients, i.e. shear strains.

The RS Lagrangian density, \cref{eqn:continuumLagrange}, is expressed as:
\begin{widetext}
\begin{align}
    \mathcal{L} & = \underbrace{\frac{\rho}{2} |\dot{u}|^2 - 2 G_1 \left (\mathrm{Re}[\partial_{\bar{z}} u]\right )^2 - 2G_2 \left (\mathrm{Im}[\partial_{\bar{z}} u] \right )^2}_{\mathcal{L}_\mathrm{conf}} - 2 B \left (\mathrm{Re}[\partial_z u]\right )^2 + \mathcal{O} \left (\frac{G_1}{N} \right ) \label{eqn:complexLagrange} 
\end{align}
\end{widetext}
where $\mathcal{L}_{\mathrm{conf}}$ is the conformal Lagrangian density describing an ideal-dilational metamaterial with negligible bulk modulus $B \ll G_2,G_1$ and large lattice size $N\gg 1$.

Of particular note are a special class of deformations that exhibit zero shear strain and are therefore of low energy: conformal maps  of the form $z \mapsto z + u(z)$, which by definition satisfy  the Cauchy-Riemann equation $\partial_{\bar{z}} u = 0$ \cite{Muskhelishvili2010}. 
These maps preserve local shape (and angles) while inducing locally isotropic expansions and compressions in the material, i.e. purely dilational deformations \cite{michaelconformal}.
Some simple conformal maps, which locally map square patches onto rotated and dilated square patches, are shown in \FigPanels{fig:figure3}{a(i-v)}  , in contrast with a non-conformal map that shears square patches into parallelograms, e.g. \FigPanels{fig:figure3}{a(vi)}.
For ideal-dilational systems governed by $\mathcal{L}_{\mathrm{conf}}$ only, these deformations cost zero energy and generate no return force.

In contrast, for realistic systems with the full dynamics, the finite terms of $\mathcal{O}(B)$ and $\mathcal{O}(G_1/N)$ in \cref{eqn:complexLagrange} may be regarded as perturbations on these otherwise zero-frequency conformal modes.
We project these full dynamics into the space of conformal functions consistent with the system's clamped corners, resulting in the predicted low-frequency conformal modes shown in \FigPanels{fig:figure4}{a}, which show good agreement with the low-frequency normal modes of the full numerical linear discrete model.
At higher frequencies, the modes become more non-uniform but nevertheless retain a characteristic feature of conformal maps: the maximum displacements (and dilations) occur exclusively at the boundary of the domain, due to the maximum modulus principle \cite{Conway_1978}.
As shown in \FigPanels{fig:figure4}{b}, the dispersion of these conformal modes (solid blue line) for experimentally realistic parameters tracks closely with the direct numerical calculations using the linearised discrete model for the first few modes (black solid line), before transitioning to a bulk phonon band. Lattices of larger size and smaller bending stiffness display a larger number of conformal modes at lower frequencies with a roughly linear relation to mode number (dashed blue \& black line). The continuum conformal frequencies are above those of the discrete model due to corrections due to the clamped boundary conditions, finite-size effects and finite shear, and coincide with them when these conditions are excluded.
Our theory is in agreement with recent predictions of dispersion for isotropic conformal elasticity of large lattice size \cite{Cheng2023}, and expands it to anisotropic designs (e.g. RS metamaterial) of experimentally realisable sizes.
The bulk phonon modes (red lines) are likewise derived from the conformal Lagrangian density $\mathcal{L}_{\mathrm{conf}}$ [\cref{eqn:complexLagrange}], with speeds of propagation for the plane wave solutions given by $c_{1,2} = \sqrt{G_{1,2}/\rho}$, correctly predicting a ceiling on the conformal band at frequency $f_{\mathrm{ceil}} (\mathrm{Hz}) \approx \sqrt{G_2/(2\rho N a^2)}$ ($\sim 31.5$ Hz for experimental parameters). More details about the non-conformal bulk band can be found in Supplementary Material Section VI S3.

To quantify the degree of conformal nature in the normal modes of the linear model, we measure the ratio of spatially averaged shear to total strain, $\langle \Psi \rangle_s$, same as presented earlier in \FigPanels{fig:figure1}{d} but without any intrinsic time dependence for mode profiles, as well as the fractional deviation of the actual displacement profile from the nearest conformal map drawn from a finite-order polynomial basis, as described in the Supplementary Information Section VII S1.
As shown in \FigPanels{fig:figure4}{c}, below the frequency ceiling, both measures show that the conformal model accurately captures the response, with as much as $90\%$  of the mode captured by the conformal fit. 
As parameters are shifted beyond those currently experimentally achievable, more modes become nearly conformal, with some more than $99\%$ conformal. \FigPanels{fig:figure4}{d} shows the effect of reducing the bending stiffness of the hinges, and by extension lowering the effective bulk modulus of the system, which improves the conformal fit of the first few modes markedly before saturating. As shown in \FigPanels{fig:figure4}{e}, this saturation can be lowered by increasing the number of blocks in the system, reflecting that the deviations from ideal conformal mechanics come both from finite bending stiffness and finite-size effects.

Despite such complications, the response of the periodically driven $30^{\circ}$ and $5^{\circ}$ experimental systems tracks closely with both the aforementioned linear and nonlinear discrete models, as shown in Fig.~\ref{fig:figure5}. Up to the predicted frequency ceiling of $31.5$ Hz, the experimental conformal fit parameter $\Delta_\textrm{conf}$ plotted in \FigPanels{fig:figure5}{a} matches closely with the model predictions and even improves upon them. The non-conformal deviations go below $10\%$ for the $30^\circ$ sample when driven near the first conformal mode, with the corresponding dynamic fit to conformal maps shown in Supplementary Video 1. When the response of that system is compared in \FigPanels{fig:figure5}{b} to the first four predicted modes, which include small non-conformal corrections, the agreement is even sharper with up to $97\%$ of the response accounted for. \FigPanels{fig:figure5}{c} shows the spatial alignment between an experiment snapshot and the first conformal mode.

\begin{figure*}
\centering
    \begin{subfigure}[t]{\linewidth}
        \includegraphics[width=\textwidth]{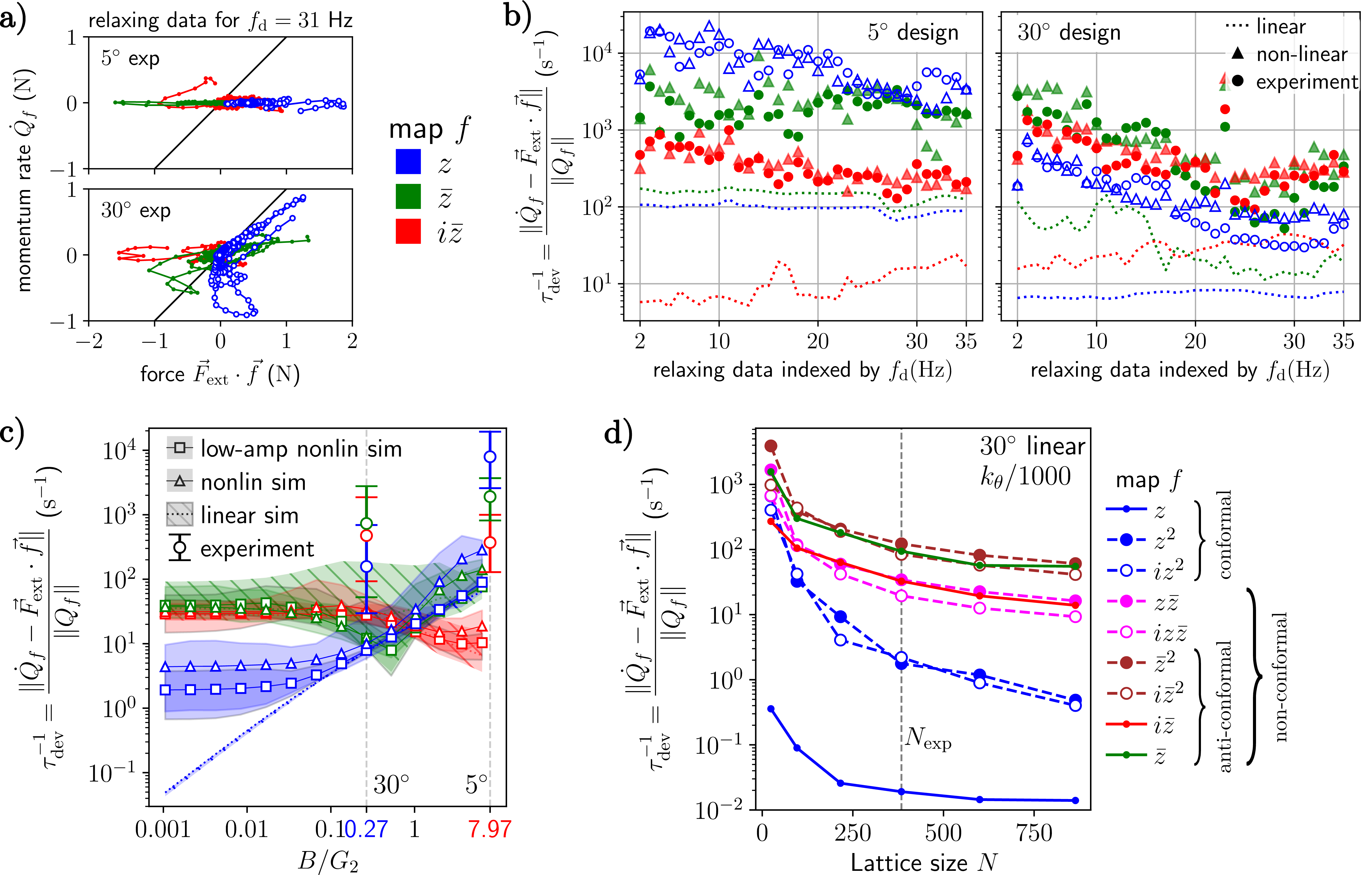}
    \end{subfigure}
\caption{
\textbf{(a)} Momentum rates vs external forces along complex monomials, conformal $z$ (dilational momentum) and non-conformal $\bar{z}$ and $i\bar{z}$ (shear momenta), for relaxing regime of experiments conducted at drive frequency $31$ Hz. For $30^{\circ}$ sample, the conformal z-momentum (blue) lies close to $y=x$ line (black) of ideal conservation compared to the other two momenta, whereas this is not the case for $5^{\circ}$ sample.
\textbf{(b)} Plots show the deviation from exact conservation for the same three momenta in relaxing regime, represented as an inverse timescale of momentum change due to internal forces ($\tau_{\mathrm{dev}}^{-1}$), for experimental and simulated (linear \& nonlinear) datasets indexed by drive-frequency $f_{\mathrm{d}}$. For $30^{\circ}$ sample, these normalised deviations are lower for conformal z-momentum than the non-conformal shear momenta, whereas that isn't the case in the $5^{\circ}$ sample, with shear-2 momentum having the lowest deviations.
\textbf{(c)} Simulations for $30^{\circ}$ design with decreasing $B/G_2$ (and bending dissipation) show reduced deviations from exact conservation for the conformal $z$-momentum. Nonlinear simulations show saturation, with lower amplitudes ($A=0.01~\mu\text{m}$) giving a slight improvement over experimental amplitude, while linear estimates scale down linearly with $B/G_2$. Nonlinear calculations in this plot use a relaxing response after an ideal 10-cycle sinusoidal drive, rather than the experimentally measured realization of the same protocol. Shaded regions \& error bars represent range of $\tau^{-1}_{\mathrm{dev}}$ values across different $f_{\mathrm{d}}$, whereas bold markers represent the mean.
\textbf{(d)} Linear simulations with varying lattice size $N$ for a more dilational case ($30^{\circ}$, $k_{\theta}/1000$) show drastic improvement in higher-order conformal momenta ($z^2$, $iz^2$) towards exact conservation compared to non-conformal momenta of same order, since finite-size effects decrease for larger $N$.
}
\label{fig:figure6}
\end{figure*}

\section{\label{sec6} Conformal Symmetry}

While conformal boundary modes are present at low frequencies, the dynamics at high frequencies and in the bulk are strongly influenced by an approximate conformal symmetry.
This continuous symmetry corresponds to transforming any trajectory, $u(z,\bar{z},t)$, by addition of an infinitesimal static conformal deformation $f(z)$ as illustrated in \FigPanels{fig:figure3}{b}.
The added deformation does not generate any potential or kinetic energy for the ideal-dilational limit, $B \to 0$ and $1/N \to 0$, leaving the conformal Lagrangian density $\mathcal{L}_{\mathrm{conf}}$ [\cref{eqn:complexLagrange}] invariant.
Hence, the dynamics remain unchanged under any such arbitrary conformal mapping of the system, defining a unique continuous symmetry corresponding to each independent conformal generator.

As established by Noether's Theorem \cite{Noether1918, spontaneousbreak}, any continuous symmetry, such as translational invariance, imposes a strong restriction on the possible trajectories a system can follow in the form of a local conservation law. Canonical examples are the conservation of linear and angular momenta implied by continuous symmetries of an isolated system corresponding global (uniform) translations and rotations, respectively.
Similarly, we find a unique conservation law corresponding each conformal-symmetry generator $f(z)$ that restricts the dynamics of an ideal-dilational material through a local continuity equation for the momentum density along $f$, $p_f(z,\bar{z},t):=\mathrm{Re}[\rho \dot{u}\bar{f}]$.
A few simple conformal generators are shown in \FigPanels{fig:figure3}{a(i-v)}.

Real dilational systems [\cref{eqn:complexLagrange}] explicitly, but weakly, break this conformal symmetry due to perturbative terms.
Given system size $L$, for a conformal generator $f(z)$, the Lagrangian density changes by $\mathcal{O}(f' B) + \mathcal{O}(L f'' G_1/N)$, indicating the dynamics remain approximately invariant if $B$ is small and $N$ is large, for a deformation $f(z)$ that varies gradually compared to unit cell size.
Under such weak symmetry breaking, we find an approximate continuity equation for momentum density $p_f$,
\begin{align}
    \partial_t p_f + \partial_z j_f + \partial_{\bar{z}} \overline{j_f} & = \mathcal{O}(f'B) + \mathcal{O}(L f'' G_1/N) \label{eqn:continuity}
\end{align}
where $\partial_z j_f + \partial_{\bar{z}} \overline{j_f}$ is the divergence of the complex current density field, $j_f(z,\bar{z},t)$, that captures the net conformal momentum flowing out of a local patch. 
This current density is related to the local stress field acting along $f$,
\begin{align}
    j_f = -(G_1 s_1 + iG_2 s_2) \overline{f} + \mathcal{O}(f B) + \mathcal{O}(f G_1/N) \label{eqn:complexcurrent}
\end{align}
where $G_1 s_1 + iG_2 s_2$ is the  complex shear stress and $\bar{f}$ is complex conjugate of $f$.

The total conformal momentum along $f$, $Q_f(t) := \int_D p_f d^2z$, follows an approximate global conservation law:
\begin{align}
    \dot{Q}_f(t) -\int_{\partial D} \mathrm{Im}[j_f d\bar{z}] & = \int_D \mathcal{O}(f'B) + \mathcal{O}(L f'' G_1/N) \, d^2z \label{eqn:globalcons}
\end{align}
where the right hand side goes to zero for ideal designs.
The boundary integral $\int_{\partial D} \mathrm{Im}[j_f d\bar{z}]$ is the net external force acting along map $f$ that balances the internal stresses at the boundary, and accounts for the net influx of momentum $Q_f$ from external sources. In contrast, the area integral accounts for momentum change due to internal forces that originate from the symmetry-breaking terms in the Lagrangian. These cause small deviations in the conservation of conformal momentum $Q_f$ that scale with bulk modulus $B$, inverse system size $1/N$ and the map's non-uniformity $\|f'\|$. Hence, in the ideal-dilational limit, these conformal momenta become constants of motion of the system, changing only due to flow of momentum from external sources and sinks.

We now apply this conserved momentum model to the behaviour of experimental systems undergoing relaxation following a drive at $31$ Hz.
The external momentum input along arbitrary complex map $f(z,\bar{z})$ is estimated using the external forces $\Vec{F}_{\mathrm{ext}}$ supplied by the dynamical boundary constraints as $\int_{\partial D} \mathrm{Im}[j_f d\bar{z}] \approx \Vec{F}_{\mathrm{ext}} \cdot \Vec{f}$, where $\vec{f}$ denotes the movement of all blocks in the lattice as defined by $f$. This is compared to the observed change in the momentum $Q_f$ along uniform maps: conformal $z$, and non-conformal $\bar{z}$ and $i\bar{z}$.
As shown in \FigPanels{fig:figure6}{a}, the (non-conformal) $5^{\circ}$ sample freely violates this law, whereas the $30^{\circ}$ sample shows strong agreement between the external force and conformal momentum associated with uniform dilation $z$ (blue).

Note that we only analyse the relaxation data here from experiments and simulations when the drive is inactive because an active driving introduces an external timescale (drive frequency $f_{\mathrm{d}}$) that controls the temporal variance in any momentum, irrespective of it being conformal or non-conformal. This timescale decides the dominant balance of terms in \cref{eqn:globalcons}, since different terms scale differently with drive frequency.

We quantify the deviation from conservation of a momentum indexed by map $f$ by measuring the discrepancy between its rate of change and the external force along its map, normalized by the momentum magnitude. This measure gives the inverse of a deviation timescale for variance in $Q_f$ unexplained by external forces:
\begin{align}
    \tau_{\mathrm{dev}}^{-1} & := \frac{\|\dot{Q}_f - \vec{F}_{\mathrm{ext}} \cdot \vec{f} \|}{\| Q_f \|}.
\end{align}
For the relaxing responses of the non-dilational $5^\circ$ sample following driving at different frequencies, \FigPanels{fig:figure6}{b} shows that the dilational momentum ($Q_z$, blue) is multiple times \emph{less} conserved than momenta corresponding to uniform shear maps (green and red). In contrast, for the dilational $30^\circ$ sample, $Q_z$ is the most conserved momentum with lowest $\tau^{-1}_{\mathrm{dev}}$.
Comparing the two samples, the conformal momentum $Q_z$ varies about 100 times \emph{slower} in the $30^{\circ}$ sample, due to smaller internal forces, than in the $5^{\circ}$ sample.

The remaining discrepancy between the rate of change of $Q_z$ and corresponding external force in the conformal $30^{\circ}$ sample is attributable not just to finite-size effects and the small bulk modulus, but also to dissipation and nonlinearities.
Hence, as shown in \FigPanels{fig:figure6}{c}, nonlinear simulations lacking significant dissipation confirm that this conservation law strengthens as the system becomes more dilational, with linear simulations showing even smaller deviations that linearly decrease with $B/G_2$. \FigPanels{fig:figure6}{d} shows, for small $B$ and small dissipation, how conformal momenta along non-uniform maps, $z^2, iz^2$ (blue dashed), become more conserved with increasing lattice size $N$ due to weaker finite-size effects. In comparison, non-conformal momenta along uniform and non-uniform maps (solid -- red/green and dashed -- pink/brown) remain non-conserved at higher $N$.

\section{\label{concl} Conclusion}

We have investigated the dynamics of two-dimensional dilational systems via the canonical model structure of rotating squares. Simulation and fabrication of two samples with differing initial angle $\theta_0$ reveal conformal behaviour in one sample ($30^{\circ}$) and non-conformal behaviour in the other ($5^{\circ}$) when dynamically excited. In the $30^{\circ}$ sample, we find that the lowest frequencies of excitation activate conformal boundary modes, consisting of characteristic patterns of dilation and rotation. At higher frequencies, bulk modes contain significant shear and are thus not themselves conformal, but demonstrate conservation of novel physical quantities which we term conformal momenta. These quantities are analogous to conventional linear momenta, but are approximately conserved  due to the approximate conformal symmetry of the structure. Numerical simulations demonstrate that further experimental refinements can enhance both of these effects by lowering the effective bulk modulus and reducing finite-size effects. Conserved quantities play key roles in fundamental physics and in the integrability of dynamical systems, in which they permit quantitative predictions of nonlinear deformations such as solitons, which have also been demonstrated in such systems~\cite{Deng2017,Deng2018b,Deng2019b}. The conformal elasticity demonstrated here may give rise to new methods of control and prediction for complex, nonlinear flexible structures, as well as for novel methods of mechanical signal processing and energy transfer \citep{WU2021,Vakakis2022}.
This conformal model can be applied to a growing class of 2D dilational designs \citep{Cho2014,Acuna2022,peng2026} even at micron scale \citep{Melio2026}, and possibly expanded to include even 3D dilational structures \citep{Buckmann2014}.

In addition to conformal symmetry, there are related flexible structures, composed of corner-sharing non-square quadrilaterals or so-called planar kirigami, which have their own modes~\cite{ian_paul_conformal,czajkowski2022duality}. Depending upon whether the deformation mechanism is auxetic or anauxetic, the analogous low-frequency modes may exist in either the boundary or the bulk, and conserved quantities may be likewise qualitatively different in such structures.
In combination, these results demonstrate a fundamentally new framework that both grants new methods of controlling nonlinear waves and develops new ways of tying fundamental physical symmetries into the dynamics of real systems.

\begin{acknowledgments}
We gratefully acknowledge support from the ARO MURI program (W911NF-22-1-0219), and from NSF PHY-2309135 to the Kavli Institute for Theoretical Physics
(KITP) (DZR).
\end{acknowledgments}

\section*{Data Availability}
The data supporting the findings of this work will be openly available at \href{https://github.com/bertoldi-collab/dynamic-conformal-metamaterials}{github.com/bertoldi-collab/dynamic-conformal-metamaterials}

\bibliography{refs}

\end{document}


\newtheorem{theorem}{Theorem}
\newtheorem{definition}{Definition}

\newcommand{\bigzero}{\mbox{\normalfont\Large\bfseries 0}}

\newcommand{\dr}{\mathbf{r}}
\newcommand{\dq}{\mathbf{q}}
\newcommand{\du}{\mathbf{u}}
\newcommand{\de}{\mathbf{e}}
\newcommand{\dv}{\mathbf{v}}
\newcommand{\dn}{\mathbf{n}}
\newcommand{\dk}{\mathbf{k}}
\newcommand{\dx}{\mathbf{x}}
\newcommand{\dl}{\mathbf{l}}
\newcommand{\dF}{\mathbf{F}}
\newcommand{\ub}{\bar{u}}
\newcommand{\zb}{\bar{z}}
\newcommand{\fb}{\bar{f}}
\newcommand{\gb}{\bar{g}}
\newcommand{\kb}{\bar{k}}

\renewcommand\thefigure{S\arabic{figure}} 
\renewcommand\thepage{S\arabic{page}}
\renewcommand\thesubsection{S\arabic{subsection}}
\renewcommand\thesubsubsection{\thesubsection.\arabic{subsubsection}}
\renewcommand\theequation{S\arabic{equation}}

\makeatletter
\renewcommand{\p@subsubsection}{\thesection~} 
\makeatother

\preprint{APS/123-QED}

\title{Supplementary Information}

\author{Neel Singh}

\author{Audrey Watkins}%
\author{Giovanni Bordiga}

\author{Vincent Tournat}

\author{Katia Bertoldi}
\author{Zeb Rocklin}

\date{\today}
\maketitle

\tableofcontents

\newpage 

\section{Fabrication}
To physically realize our mechanical metamaterials, we employ a fabrication process that combines 3D printing with customized assembly~\cite{Bordiga2024}. 

\subsection{Shim Laser Cutting}
Long rectangular strips of width \SI{6.35}{mm} are laser cut from commercially available green polyester plastic sheets of thickness \SI{76}{\um} (McMaster, part number 9513K65). These shims will act as hinges and couple neighboring blocks to each other.

\subsection{CAD Design}

We import sketches of the periodic lattice geometries into a CAD software (onShape), where the 2D components are represented as blocks and hinges. The hinges are specifically designed to create a continuous path, which plays a crucial role in guiding the placement of the polyester shims later in the fabrication process. The block sketches are extruded to a height of 8.1 mm, with a rectangular cross-section exclusion (6.35 mm in height and 0.5 mm in width) integrated along the continuous hinge pathway to create space for the insertion of polyester shims. The generated CAD files are then imported to the open-source slicer software (Cura) and prepared for 3D printing. All samples are printed using white polylactic acid (PLA) filament on an UltiMaker\textsuperscript{\textregistered} 3 3D printer.

\subsection{3D Printing, Customized Assembly, and Post-Printing Processing}

The 3D printing G-Code for all fabricated samples is modified to pause at a height of 7.2 mm to allow for the manual insertion of the laser-cut polyester shims into their exclusions. To ensure adhesion between the shims and 3D printed blocks, a small amount of glue (cyanoacrylate glue) is applied to the internal corners of the shim exclusions. After the applied glue dries, the 3D printing is resumed, and the shims are fully encapsulated by the PLA.

To remove any residual stresses introduced to the shims during the customized assembly process, the entire glass print bed and the attached 3D printed structure are placed in an 80 $^\circ$C oven for 60 minutes. After heating, the sample cools to room temperature and is then removed from the glass print bed. This heating process ensures that the geometry of the fabricated sample is not altered due to residual stresses.

\section{Dynamic Testing}

Our experimental setup consists of a high-speed camera (Phantom$^{\text{®}}$ Miro C211), low-frequency shaker (The Modal Shop$^{\text{©}}$ model 2025E), waveform function generator (Tektronix$^{\text{®}}$ AFG 31000), and amplifier (APS Dynamics$^{\text{©}}$ PA 500 DM).

\begin{figure*}[h!]
    \includegraphics[width=\textwidth]{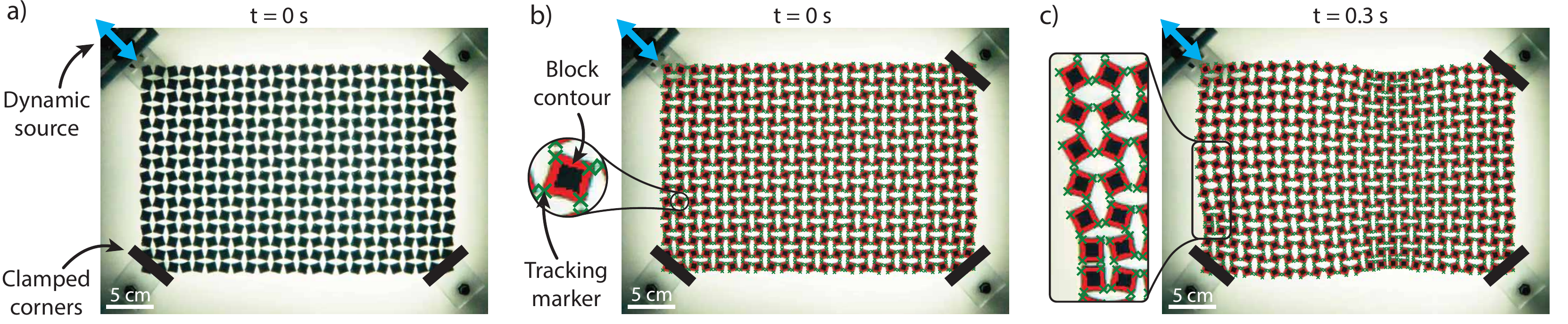}
    \caption{\textbf{Digital Image Correlation.} \textbf{(a)} Snapshot of the first frame at $t=0$ s of the $\theta_0=30^{\circ}$ sample. \textbf{(b)} Snapshot of the first frame at $t=0$ s with the block contours (red) and tracking markers (green) overlaid on each block. \textbf{(c)} Snapshot at $t=0.3$ s capturing the dynamic response of the sample with block contours and tracking markers overlaid at their correct spatial locations.}
\label{fig:SI_digital_image_correlation}

\end{figure*}

Three of the four corners of each fabricated sample are suspended and bolted to an optical table using optical posts [\Cref{fig:SI_digital_image_correlation}a]. The fourth corner is attached to the low-frequency shaker using a custom 3D printed press-fit attachment. The low-frequency shaker drives the displacement of the fourth corner with a dynamic signal created and amplified by the function generator and amplifier, respectively. The high-speed camera is mounted above the sample and is connected to the function generator to ensure time synchronization between dynamic excitation and video recording. LED light fixtures and a light diffusing plate are placed below the suspended sample to ensure high-contrast video recordings for post-processing. We trigger the dynamic input and the camera captures the response of the sample, recording at a frame rate of 1000 frames per second.

To quantify the experimental dynamic response of our samples, we use a custom Digital Image Correlation software based on the OpenCV open-source computer vision library to track the translational and rotational displacements of all rigid blocks. The first frame [\Cref{fig:SI_digital_image_correlation}a] of each high-speed experimental video undergoes grayscale thresholding to capture the contour of each rigid PLA block. Tracking markers are then placed at all four corners of each block contour [\Cref{fig:SI_digital_image_correlation}b]. The displacement of each marker between consecutive frames is determined by using a cross-correlation within a small local search window, allowing the contours and markers to update with each frame [\Cref{fig:SI_digital_image_correlation}c]. This process is repeated in each video frame and for all markers. Once all video frames are tracked, the translational and rotational displacements of the block centroids are determined from the history of the tracking marker positions.

\newpage

\section{Discrete Nonlinear Dilation and Shear Calculations}
\label{section: SI discrete nonlinear dilation and shear}

\begin{figure*}[h]
\centering
    \includegraphics[width=0.6\textwidth]{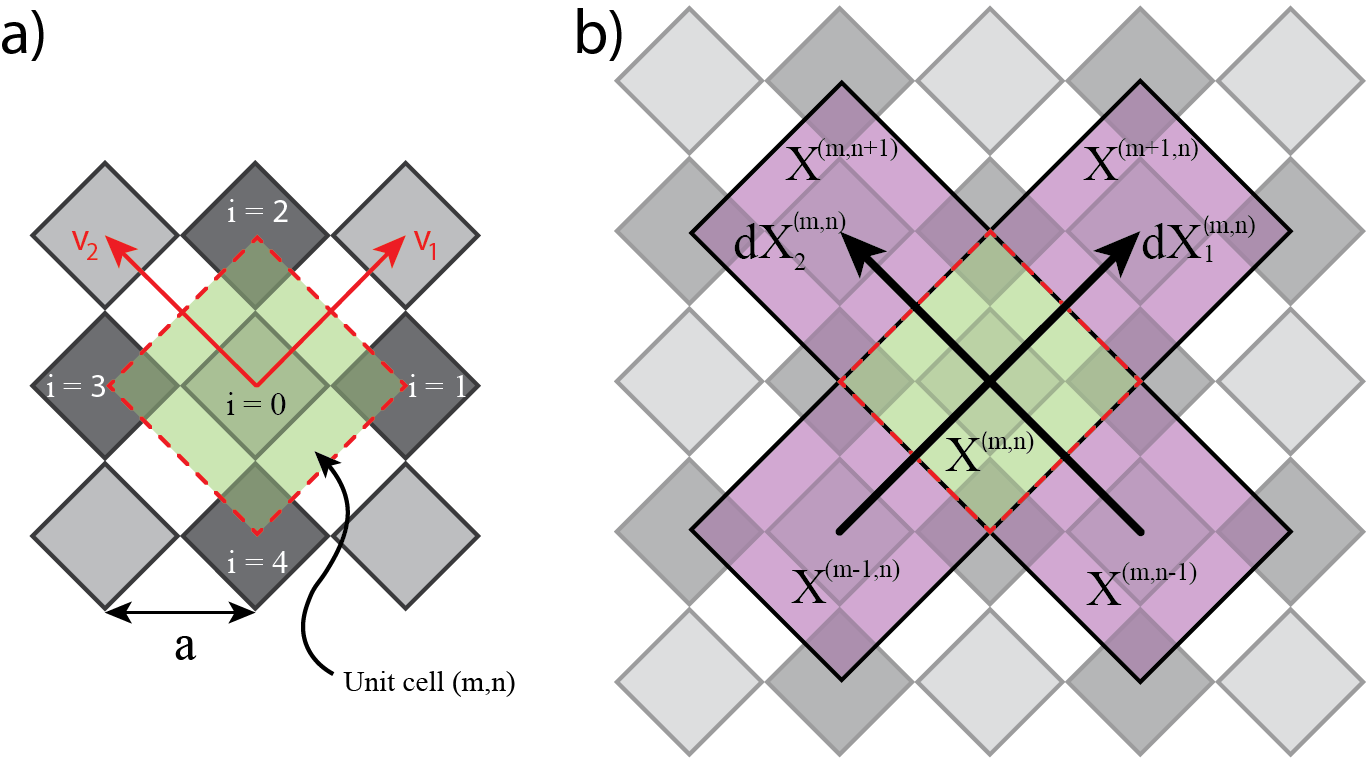}
    \caption{\textbf{Unit cell tessellation.} a) Schematic of 3 $\times$ 3 rotating square mechanism with unit cell ($m,n$) and its basis vectors highlighted in red, along with the indices of each block included in the calculation for unit cell position $\textbf{x}^{(m,n)}$. b) Schematic of a 5 $\times$ 5 rotating square mechanism with unit cell ($m,n$) highlighted in green and the surrounding unit cells used to calculate dilation, $d^{(m,n)}$, and shear, $s^{(m,n)}$, magnitudes highlighted in purple.}
    \label{fig:SI_fig_deformation}
\end{figure*}

\Cref{fig:SI_fig_deformation}a shows the unit cell and basis vectors of our mechanical metamaterial consisting of periodic rotating squares with center-to-center distance $a$. The unit cell consists of one entire central block, indexed as $i=0$, and a quarter of its four surrounding adjacent blocks, indexed as $i = 1, 2, 3, 4$. The unit cell is tessellated using the basis vectors $\mathbf{v}_1 = a ( \hat{x} + \hat{y})$ and $\mathbf{v}_2 = a (-\hat{x} + \hat{y})$, where  $\hat{x}$ and $\hat{y}$ are unit vectors aligned with the $x$ and $y$ directions, respectively. We define the position of unit cell ($m,n$) as a weighted average of the displacements of all block centroids in the unit cell and calculate the current position at time $t$ as 

\begin{equation}
    \textbf{x}^{(m,n)}(t) = \textbf{X}^{(m,n)} + \frac{1}{8}(4\textbf{u}_0^{(m,n)}(t) + \textbf{u}_1^{(m,n)}(t) + \textbf{u}_2^{(m,n)}(t) + \textbf{u}_3^{(m,n)}(t) + \textbf{u}_4^{(m,n)}(t)) ,
\end{equation}
where $\textbf{X}^{(m,n)}$ is the position of the centroid of the unit cell at $t=0$ s, and $\textbf{u}_i(t)^{(m,n)}$ with $i \in [0, 4]$ is the displacement of the centroid of each of the five blocks accounted for in the unit cell [\Cref{fig:SI_fig_deformation}a].

To calculate the nonlinear dilational and shear strain magnitudes   of unit cell ($m,n$), we consider the positions of the four surrounding unit cells as shown in \Cref{fig:SI_fig_deformation}b and define its deformation gradient as
\begin{equation}
    \textbf{F}^{(m,n)} = \text{d}\textbf{x}^{(m,n)}\left(\text{d}\textbf{X}^{(m,n)}\right)^{-1},
\end{equation}
where
\begin{align}
    \text{d}\textbf{X}^{(m,n)} = (\text{d}\textbf{X}_{1}^{(m,n)}, \text{d}\textbf{X}_{2}^{(m,n)}) &= (\textbf{X}^{(m+1,n)} - \textbf{X}^{(m-1,n)}, \textbf{X}^{(m,n+1)} - \textbf{X}^{(m,n-1)}) \\
    \text{d}\textbf{x}^{(m,n)}=(\text{d}\textbf{x}_{1}^{(m,n)}, \text{d}\textbf{x}_{2}^{(m,n)}) &= (\textbf{x}^{(m+1,n)} - \textbf{x}^{(m-1,n)}, \textbf{x}^{(m,n+1)} - \textbf{x}^{(m,n-1)}).
\end{align}
Finally, we define the nonlinear dilation and shear magnitudes as
\begin{equation}
    d^{(m,n)} = \sqrt{\text{det}[\textbf{C}^{(m,n)}]} -1,
    \label{SIeqn:dilation}
\end{equation}
and
\begin{equation}
    \left(s^{(m,n)}\right)^2 = \text{tr[\textbf{C}$^{(m,n)}$]} - 2\sqrt{\text{det[\textbf{C}$^{(m,n)}$]}} ,
    \label{SIeqn:shear}
\end{equation}
where $\textbf{C}=\textbf{F}^T\textbf{F}$ is the  right Cauchy-Green deformation tensor. 

\newpage

\section{Discrete Model for RS Metamaterial}

\begin{figure}[h!]
\centering
    \begin{subfigure}[t]{0.8\linewidth}
        \includegraphics[width=\textwidth]{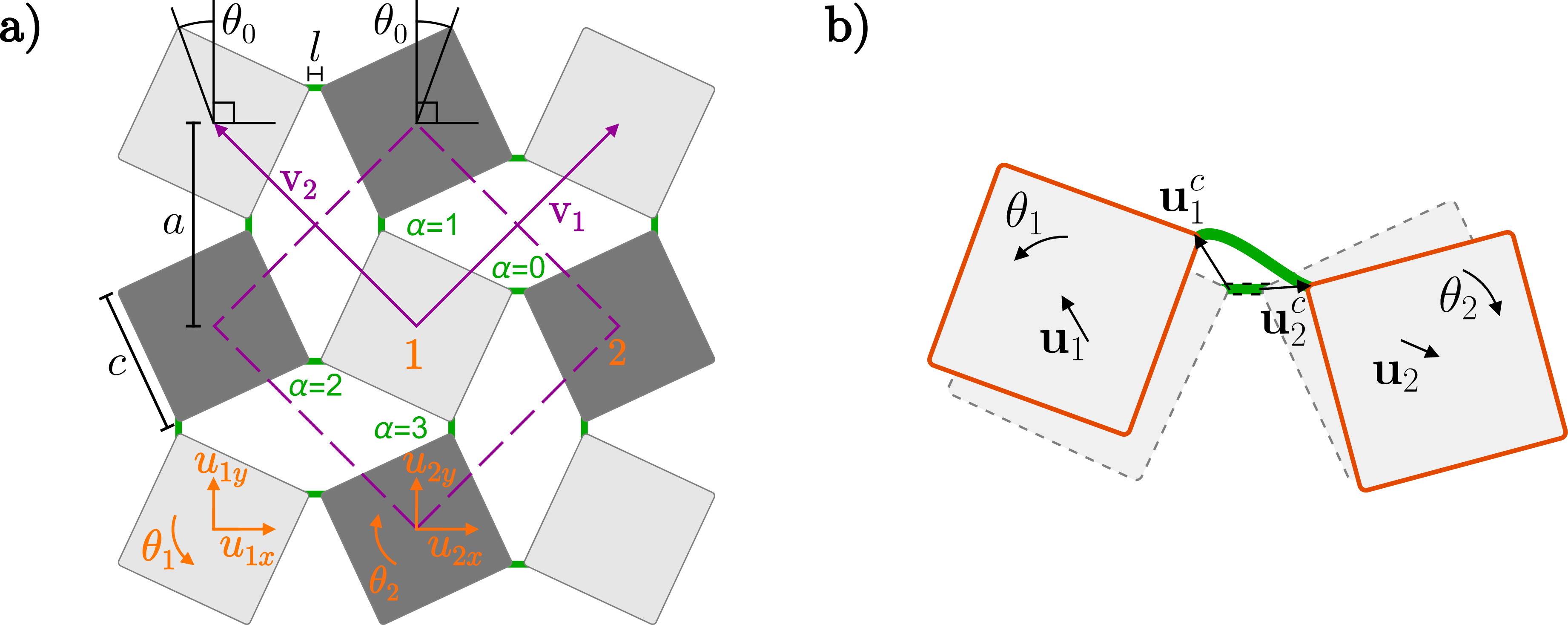}
    \end{subfigure}
\caption{\textbf{(a)} The reference state of the RS lattice can be defined by a unit cell, consisting of two squares (indexed $1$ and $2$) rotated in-plane in opposite directions by the same angle $\theta_0$ in reference to the fully expanded state, repeated at integer multiples of lattice vectors $\dv_1$ and $\dv_2$, with every square corner indexed by $\alpha$, and connected by a thin ligament (shown in green) to its nearest neighbour's corner. The spacing between adjacent squares is related to the square side-length $c$, ligament length $l$ and orientation $\theta_0$ as $a = \sqrt{2}c\cos{\theta_0} + l$. \textbf{(b)} The degrees of freedom for a unit cell are given by the block (square) centre-of-mass displacements $\du_1$ and $\du_2$, and their in-plane rotations $\theta_1$ and $\theta_2$, corresponding to, respectively, squares $1$ and $2$, giving the six scalar degrees of freedom per unit cell. The rotations are measured in an alternating convention, such that $\theta_1$ is counter-clockwise rotation of square $1$ and $\theta_2$ is the clockwise rotation of square $2$ from the reference orientations.}
\label{fig:SI_lattice_dofs}
\end{figure}

To numerically model the dynamics, we choose the squares $1$ and $2$ as shown in \Cref{fig:SI_lattice_dofs}a to compose our repeating lattice, such that square $1$ is the centre and square $2$ in right corner of the unit cell.
There are six degrees of freedom per unit cell, given by the block-centroid displacements $\du_1 = u_{1x} \hat{x} + u_{1y} \hat{y},\, \du_2 = u_{2x} \hat{x} + u_{2y} \hat{y}$, and the in-plane block rotations $\theta_1, \theta_2$, as shown in \Cref{fig:SI_lattice_dofs}b. The rotations are defined in an alternating convention, i.e. $\theta_1$ is measured counter-clockwise and $\theta_2$ is measured clockwise, so that the rotations associated with dilational mechanism are uniform: $\theta_1 = \theta_2$. While the blocks are rigid, their movements can deform the connecting ligaments (hinges).

\subsection{Hinge Energetics}
\label{subsec:hinge_symm_deform}

\begin{figure*}[h]
\centering 
    \begin{subfigure}[t]{
    \linewidth}
        \includegraphics[width=\textwidth]{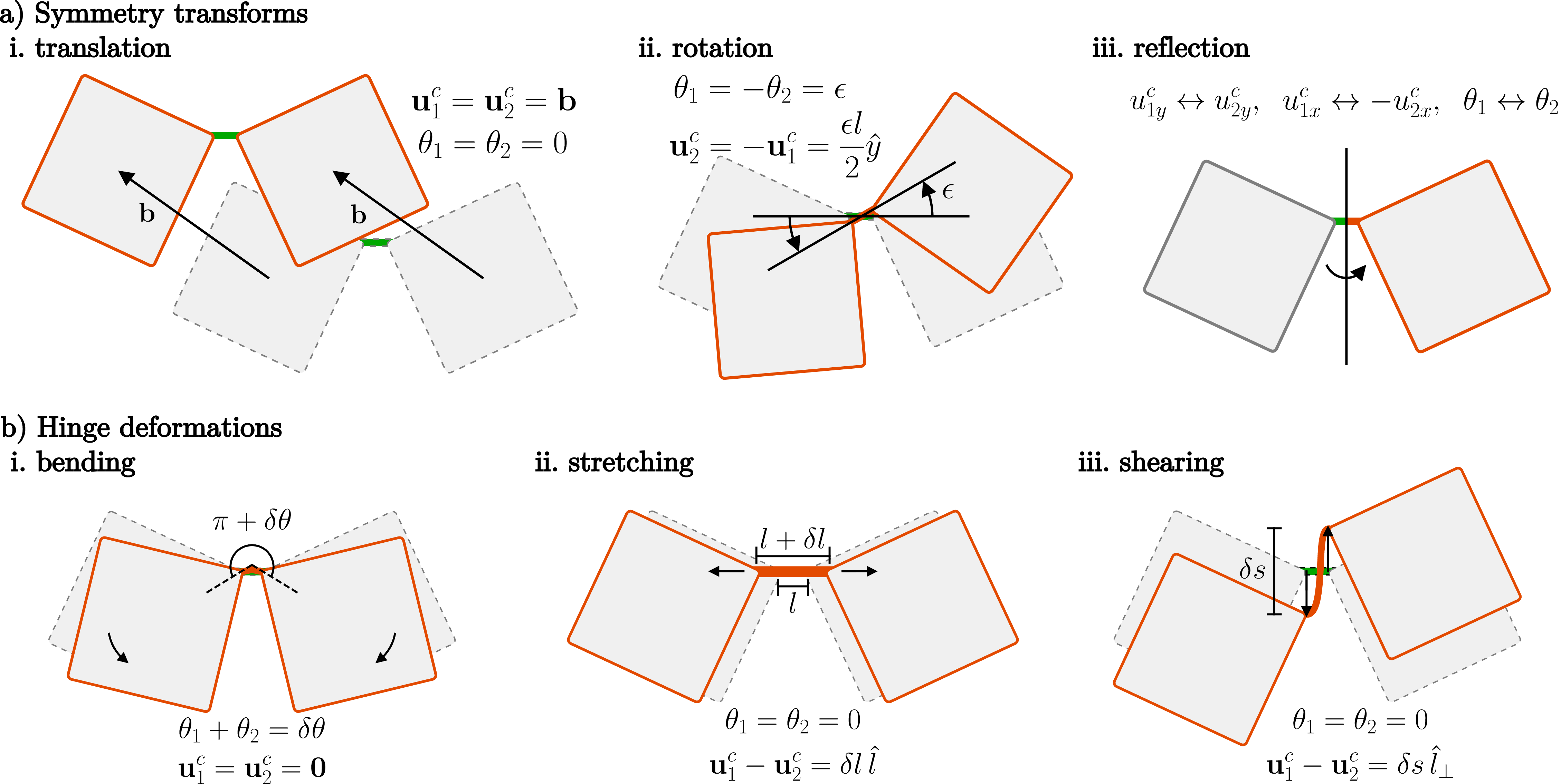}
    \end{subfigure}
\caption{\textbf{(a)} Symmetry transformations of block-ligament-block system: \textbf{i.} uniform translation, \textbf{ii.} uniform rotation, and \textbf{iii.} mirror reflection about the perpendicular bisector of ligament ($\hat{l}_{\perp}$). 
\textbf{(b)} Three independent hinge deformations in linear limit: \textbf{i.} pure bending, \textbf{ii.} pure stretching, and \textbf{iii.} pure shearing.}
\label{fig:SI_hinge_symmdeform}
\end{figure*}

The deformation of a ligament (hinge) depends directly on the displacements of connected corners, $\du_1^c$ and $\du_2^c$, and the block rotations about these corners, $\theta_{1}$ and $\theta_{2}$, shown in \Cref{fig:SI_lattice_dofs}b. These are, in turn, related to the degrees of freedom of the two rigid blocks connected by the ligament: their centroid displacements and in-plane rotations. The generalised forces (including moment forces) generated in the ligament are assumed to be linear in the corner displacements and rotations, lying in a $6$-dimensional space. Due to symmetries of the two-block system, shown in \Cref{fig:SI_hinge_symmdeform}a, we can ascertain that some combinations of corner movement (and, in turn, block movements) cannot deform the ligaments.

Translational symmetry of the system implies that any overall translation of the two neighbouring blocks does not deform the ligament (see \Cref{fig:SI_hinge_symmdeform}a i). Therefore, only the relative corner displacement $\du_2^c - \du_1^c$ may deform the ligament and generate a restoring force, whereas the displacement combination $\du_2^c + \du_1^c$ does not deform the ligament.
Similarly, due to rotational symmetry (see \Cref{fig:SI_hinge_symmdeform}a ii), an overall rotation by small angle $\epsilon$ does not deform the ligament. Under this rotation, the corners displace relative to each other by $\du_{2}^c - \du_{1}^c = \epsilon l \hat{l}_{\perp} $, and rotate by $\theta_1 = -\theta_2 = \epsilon$ in our alternating convention, where $\hat{l}_{\perp}$ is the direction perpendicular to the reference ligament axis $\hat{l}$. This implies that the rotation-free linear combination $(\du_{2}^c - \du_{1}^c)\cdot \hat{l}_{\perp} + (\theta_2 - \theta_1)l/2$ generates restoring forces, but the purely rotational combination $(\du_{2}^c - \du_{1}^c)\cdot \hat{l}_{\perp} - (\theta_2 - \theta_1) l/2$ does not. Because of these symmetries, the six-dimensional space spanned by translations and rotations has three rigid-body motions that leave the ligament unstrained, and three motions that deform it and generate restoring forces. These latter modes are spanned by the following three degrees of freedom (see \Cref{fig:SI_hinge_symmdeform}b):
\begin{align}
    \left \{ \theta_1 + \theta_2, \quad (\du_{2}^c - \du_{1}^c) \cdot \hat{l} , \quad (\du_{2}^c - \du_{1}^c) \cdot \hat{l}_{\perp} + \frac{l}{2}(\theta_2 - \theta_1) \right \} .\label{SIeqn:3springext}
\end{align}

Notable recent works \cite{Coulais2017,Bordiga2024}, employing a similar discretised spring-mass model to study nonlinear phenomena in RS-metamaterial, consider the following three independent ligament deformations: bending $\delta \theta$, (longitudinal) stretching $\delta l$ and (transverse) shearing $\delta s$. In the linear limit, these choices agree with our symmetry-motivated basis of ligament deformations of \cref{SIeqn:3springext}:
\begin{align}
    \begin{aligned}
        \delta \theta & := \theta_1 + \theta_2 , \\
        \delta l & := |\dr_2^c - \dr_1^c| - l \approx \left ( \du_{2}^c - \du_{1}^c \right ) \cdot \hat{l} + \mathcal{O}\left(|\mathbf{u}_2^c - \mathbf{u}_1^c|^2\right) , \\
        \delta s & := l \left (\psi + \frac{\theta_2 - \theta_1}{2} \right ) \approx \left ( \du_{2}^c - \du_{1}^c \right ) \cdot \hat{l}_{\perp} + \frac{l}{2}(\theta_2 - \theta_1) + \mathcal{O}\left( |\mathbf{u}_2^c - \mathbf{u}_1^c|^2 \right), \label{SIeqn:springextensions}
    \end{aligned}
\end{align}
where $\dr_1^c, \dr_2^c$ are the connected-corners positions, and $\psi$ is the angle the ligament makes with its reference axis $\hat{l}$. The bending $\delta \theta$ (in radians) is dimensionless, unlike shearing and stretching extensions which have dimensions of length.

The generalised restoring forces, acting on square $1$, generated by a deformed ligament can be expressed in terms of two force components, $f_l$ and $f_s$, and a moment force $\tau$. The force components $f_l$ and $f_s$ correspond to forces longitudinal and transverse to $\hat{l}$, respectively, and the moment $\tau$ is the counter-clockwise torque applied on square $1$ from the ligament. In the linear limit, the generalised forces are related to hinge deformations by the stiffness matrix $K$:
\begin{align}
    K & = \begin{bmatrix}
        k_\theta & k_{\theta l} & k_{\theta s} \\ k_{\theta l} & k_l & k_{sl}  \\ k_{\theta s} & k_{sl} & k_s
    \end{bmatrix}  & \text{and} &&  \begin{bmatrix}
        \tau \\ f_l \\ f_s
    \end{bmatrix} & = - K \begin{bmatrix}
        \delta \theta \\ \delta l \\ \delta s
    \end{bmatrix},
\end{align}
where the stiffness parameters do not all have the same dimensional units.

The stiffness matrix can be simplified using the mirror symmetry about the perpendicular-bisector axis of the ligament, as shown in \Cref{fig:SI_hinge_symmdeform}a iii. Under the action of the mirror operation $M$, only the shearing deformation of the three hinge deformations flips its sign, whereas the other two remain the same. And similarly, the only the transverse force $f_s$ flips its sign, whereas the others remain unchanged:
\begin{align}
    M \begin{bmatrix}
        \delta \theta \\ \delta l \\ \delta s
    \end{bmatrix} & = \begin{bmatrix}
        \delta \theta \\ \delta l \\ -\delta s
    \end{bmatrix} & \text{and} & & M \begin{bmatrix}
        \tau \\ f_l \\ f_s
    \end{bmatrix} & = \begin{bmatrix}
        \tau \\ f_l \\ -f_s
    \end{bmatrix} & \text{where} &&
    M & = \begin{bmatrix}
        1 & 0 & 0 \\
        0 & 1 & 0 \\
        0 & 0 & -1
    \end{bmatrix}.
\end{align}
Since pure bending and pure stretching are symmetric eigenvectors of $M$, whereas pure shearing is an anti-symmetric eigenvector of $M$, the mirror symmetry of the system, $MKM=K$, implies there cannot be any coupling between the symmetric and anti-symmetric sectors of the hinge-deformation space, i.e. $k_{\theta s} = 0 $.

We neglect the coupling between the bending and stretching motion of a hinge ($k_{\theta l} = 0$), knowing that multiple studies \cite{Coulais2017,Bordiga2024,Deng2017,Deng2018a,Deng2019a,Deng2020} have shown that simulations with such a simplified model are good predictors for dynamic experiments of RS metamaterial. As a consequence, the resulting stiffness matrix is diagonal in the three hinge deformations,
\begin{align}
    K & = \begin{bmatrix}
        k_{\theta} & 0 & 0 \\
        0 & k_{l} & 0 \\
        0 & 0 & k_{s}
    \end{bmatrix},
\end{align}
with the deformation energy given by
\begin{align}
    E_{\mathrm{hinge}} & = \frac{1}{2} \left ( k_{\theta} \delta \theta^2 + k_l \delta l^2 + k_s \delta s^2 \right ).\label{SIeqn:hingeenergy}
\end{align}

Note that $k_{\theta}$ has dimensions of torque per unit angle, $[k_{\theta}]=[ML^2T^{-2}]$, whereas $k_l$ and $k_s$ have dimensions of force per unit length, $[k_l]=[k_s]=[MT^{-2}]$.
This difference in dimensions lends to a mechanical advantage for bending over shearing and stretching when the block spacing $a$ (or size $c$) is larger than ligament size $l$.
The magnitude of relative block displacement under a purely bending deformation of $\delta \theta$ is magnified by the ``lever-arm'' length of the block diagonal, i.e.
$|\du_2 - \du_1| \sim c \delta \theta \sim a \delta \theta $ when hinges are tiny. Consequently, the effective stiffness for force resisting bend-induced block displacements is proportional to $k_{\theta}/a^2$. Hence, bending can accommodate large displacements for the same energy cost as shearing or stretching of ligaments, i.e. $k_{\theta}/a^2 \ll k_s, k_l$, when block size is much larger than hinge size ($a\gg l$). This geometry-induced mechanical advantage explains why deformations that bend ligaments must be energetically favoured over those that shear or stretch the hinges.

\subsection{\label{subsubsec2} Lattice Energy}

With the above understanding of deformation energy in hinges, we formulate the energy functions for a unit cell, and then the whole lattice. We describe the reference configuration with no deformation by the repeating position vectors and orientations of two squares, chosen to be square $1$ of arbitrary unit cell and the square $2$ next to it on the right (see \Cref{fig:SI_lattice_dofs}a). With the lattice vectors $\dv_1, \dv_2$, each unit cell is indexed by a pair of integers $\dn=(n_1,n_2)$ such that its centre (square $1$) is located at $n_1\dv_1 + n_2\dv_2$. We index the corners of square $1$ by $\alpha \in \{0,1,2,3\}$ in counter-clockwise order, starting from the right-most corner as shown in \Cref{fig:SI_lattice_dofs}a. The reference positions of square centroids and corners are given by
\begin{subequations}
\begin{align}
    \dr_1^\dn & = n_1\dv_1 + n_2\dv_2,
    \\
    \dr_2^\dn & = n_1\dv_1 + n_2\dv_2 + a \hat{x},
    \\
    \dr_{i,\alpha}^{\dn} & = \dr_i^{\dn} + \frac{c}{\sqrt{2}}\hat{r}\left (\frac{\pi}{2}\alpha-(-1)^i\theta_0 \right ), \label{SIeqn:cornerrefconfig}
\end{align}
\label{SIeqn:refconfig}
\end{subequations}

\noindent where $c$ is the square size, and the unit vector $\hat{r}(\phi)$ is used to denote direction of a corner from the centroid:
\begin{align}
    \hat{r}(\phi) & = \cos{\phi} \,\hat{x} + \sin{\phi}\,\hat{y}.\label{SIeqn:polarunitvec}
\end{align}

The lattice deformation is described by the displacements and in-plane rotations of these squares about the reference state, $\du_i^\dn$ and $\theta_i^{\dn}$, $i=1,2$, which we index by the unit cell.
The deformation energy depends on the hinge deformations caused by the corner movements of squares, as discussed in the previous section [\cref{SIeqn:springextensions,SIeqn:hingeenergy}].
The displacements of corner $\alpha$ of square $i$ in unit cell $\dn$ (see \Cref{fig:SI_lattice_dofs}a) are expressed in terms of the reference positions [\cref{SIeqn:cornerrefconfig}] and displaced block configuration, $\du_i^{\dn}$ and $\theta_i^{\dn}$, to give:
\begin{align}
    &&\du_{i,\alpha}^{\dn}
    & = \du_i^{\dn} + \frac{c}{\sqrt{2}} \left [ \hat{r}\left (\frac{\pi}{2}\alpha-(-1)^i(\theta_0 + \theta_i^\dn) \right ) - \hat{r}\left (\frac{\pi}{2}\alpha-(-1)^i\theta_0 \right ) \right ]. \label{SIeqn:cornerdisp}
\end{align}

\begin{table}[b]
\caption{\label{tab:cellneighbourindex}
Indexing notation for neighbouring squares connected by hinges in cell $\dn$}
\begin{ruledtabular}
\begin{tabular}{c|c|l|c}
\textrm{corner $\alpha$ of square $1$ in cell $\dn$}&
\textrm{corner $\alpha+2$ of square $2$ }&
\textrm{in unit cell $\dn_{\alpha}$ } &
\textrm{bond direction $\hat{l}_\alpha$} \\
\colrule
 0  &  2 & $\dn$         & $\hat{x}$ \\
 1  &  3 & $\dn+(0,1)$   & $\hat{y}$ \\
 2  &  0 & $\dn+(-1,1)$  & $-\hat{x}$ \\
 3  &  1 & $\dn+(-1,0)$  & $-\hat{y}$ \\
\end{tabular}
\end{ruledtabular}
\end{table}

To find the hinge extensions in each cell, note that the corner $\alpha$ of square $1$ is connected to the corner $\alpha+2$ (mod $4$) of square $2$ in a neighbouring cell $\dn_\alpha$, as delineated in \cref{tab:cellneighbourindex}. So, we can find hinge extensions by applying \cref{SIeqn:springextensions} for each hinge connected to square $1$ of cell $\dn$: 
\begin{subequations} 
    \begin{align}
    \delta \theta_{\alpha}^{\dn} & = \theta_1^{\dn} + \theta_2^{\dn_{\alpha}} , \\
    \delta l_{\alpha}^{\dn} & = |l\hat{l}_{\alpha} + \du_{2,\alpha+2}^{\dn_{\alpha}} - \du_{1,\alpha}^{\dn}| - l, \\
    \delta s_{\alpha}^{\dn} & = l \left (\psi_{\alpha}^{\dn} + \frac{\theta_2^{\dn_{\alpha}} - \theta_1^{\dn}}{2} \right ),
    \end{align} \label{SIeqn:nonlinear_extensions}
\end{subequations}

\noindent where $\hat{l}_\alpha = \hat{r}(\alpha\, \pi/2)$ is the reference direction of the ligament at corner $\alpha$, with $\hat{r}(\phi)$ defined in \cref{SIeqn:polarunitvec}, and $\psi_{\alpha}^{\dn}$ is the angle this ligament makes with its reference direction $\hat{l}_\alpha$ in its deformed state.

Substituting \cref{SIeqn:cornerdisp} in \cref{SIeqn:nonlinear_extensions}, the hinge extensions can be expressed using block degrees of freedom, i.e. centroid displacements $\du_i^{\dn}$ and in-plane rotations $\theta_i^{\dn}$. The above equations showcase the inherent geometric nonlinearity in expressing hinge deformations in terms of corner displacements and corner movements in terms of square movements. Therefore, this discrete model framework is nonlinear in its most general application. To linear order, we approximate the hinge extensions using the block degrees of freedom as:
\begin{subequations} 
\begin{align}
    \delta \theta_{\alpha}^{\dn} & = \theta_2^{\dn_{\alpha}} + \theta_1^{\dn}  , \\
    \delta l_{\alpha}^{\dn} & \approx  \left ( \du_{2,\alpha+2}^{\dn_{\alpha}} - \du_{1,\alpha}^{\dn} \right )\cdot \hat{l}_{\alpha} + \mathcal{O}\left(u^2\right) \nonumber \\
    & \approx \left ( \du_{2}^{\dn_{\alpha}} - \du_{1}^{\dn} \right )\cdot \hat{l}_{\alpha} +  \frac{a-l}{2} \tan{\theta_0} \left (  \theta_1^{\dn} + \theta_2^{\dn_{\alpha}} \right ) + \mathcal{O}\left(u^2\right) + \mathcal{O}\left(\theta^2\right)  , \\
    \delta s_{\alpha}^{\dn} & \approx \left ( \du_{2,\alpha+2}^{\dn_{\alpha}} - \du_{1,\alpha}^{\dn} \right ) \cdot \hat{l}_{\perp, \alpha} + \frac{l}{2} \left ( \theta_2^{\dn_{\alpha}} - \theta_1^{\dn} \right ) + \mathcal{O}\left(u^2\right) \nonumber \\
    & \approx \left (  \du_{2}^{\dn_{\alpha}} - \du_{1}^{\dn} \right ) \cdot \hat{l}_{\perp,\alpha} + \frac{a}{2} \left ( \theta_2^{\dn_{\alpha}}  -  \theta_1^{\dn} \right ) + \mathcal{O}\left(u^2\right) + \mathcal{O}\left(\theta^2\right) ,
\end{align}
\label{SIeqn:cellbondextension}
\end{subequations}
where we substitute square size using $a= \sqrt{2}c \cos{\theta_0} + l$. Moving forward, we only express the linear approximations for the relevant dynamical variables to focus on the salient features of the theory, while explicitly calling back to the full nonlinear framework described by \cref{SIeqn:nonlinear_extensions,SIeqn:cornerdisp} whenever relevant. 

Each unit cell has a coordination number of four, i.e. there are four bonds (hinges/ligaments) per unit cell. Hence, the cell deformation energy is the sum of its four hinge deformation energies:
\begin{align}
    E_{\mathrm{cell}}[\dn] & = \sum_{\alpha=0}^3 
 \left [ \frac{k_\theta}{2} (\delta \theta_{\alpha}^\dn)^2 + \frac{k_l}{2} (\delta l_{\alpha}^{\dn})^2 + \frac{k_s}{2}  (\delta s_{\alpha}^\dn)^2  \right ]. \label{SIeqn:celldeformenergy}
\end{align}
By summing up these local cell energies we can estimate the total deformation energy for any set of block displacements and rotations as
\begin{align}
    E_{\mathrm{latt}} & = \sum_\dn E_{\mathrm{cell}}[\dn] = \frac{1}{2} \sum_{\dn} \sum_{\alpha=0}^3 
    \left [ k_{\theta} (\delta \theta_{\alpha}^\dn)^2 + k_l (\delta l_{\alpha}^{\dn})^2 + k_s (\delta s_{\alpha}^\dn)^2 \right] ,\label{SIeqn:latticedeformenergy}
\end{align}
where the extensions are generally nonlinear in the block degrees of freedom, as seen in \cref{SIeqn:nonlinear_extensions}.

For calculating the kinetic energy, we sum up the translational and rotational kinetic energies of the squares constituting a unit cell, i.e. one square $1$ and a quarter of the four neighbouring square $2$'s as shown in \Cref{fig:SI_lattice_dofs}a, to find the cell kinetic energy:
\begin{align}
    T_{\mathrm{cell}}[\dn] & = \frac{m}{2} \left ( \dot{\du}^{\dn}_1\right)^2 + \frac{m}{8} \sum_{\alpha=1}^4 \left ( \dot{\du}^{\dn_\alpha}_2\right)^2
        + \frac{J}{2} \left ( \dot{\theta}^{\dn}_1\right)^2 + \frac{J}{8} \sum_{\alpha=1}^4 \left ( \dot{\theta}^{\dn_\alpha}_2\right)^2.
        \label{SIeqn:cellkineticenergy}
\end{align}
The total kinetic energy of the system is this cell energy summed over all the unit cells:
\begin{align}
    T_{\mathrm{latt}} & = \sum_{\dn} \frac{m}{2} \left ( \left ( \dot{\du}^{\dn}_1\right)^2 + \frac{1}{4} \sum_{\alpha} \left ( \dot{\du}^{\dn_\alpha}_2\right)^2 \right )
    + \sum_{\dn}  \frac{J}{2} \left (  \left ( \dot{\theta}^{\dn}_1\right)^2 + \frac{1}{4} \sum_{\alpha} \left ( \dot{\theta}^{\dn_\alpha}_2\right)^2 \right ) ,
    \label{SIeqn:latticekineticenergy}
\end{align}
where $m$ is mass and $J = \frac{mc^2}{6}$ is out-of-plane moment of inertia of each square.

\subsection{Lattice Dynamics}

For $N$ blocks ($\approx N/2$ unit cells) in the lattice, we have $3N$ degrees of freedom, which we denote as a vector of generalised coordinates $\vec{q} = (q_1, q_2, \ldots, q_{3N}) =  ( u_{1,x}^{\dn}, u_{1,y}^{\dn}, \theta_{1}^{\dn}, u_{2,x}^{\dn}, u_{2,y}^{\dn}, \theta_{2}^{\dn} )_{\dn}$. Using the lattice potential and kinetic energy expressions, \cref{SIeqn:latticedeformenergy,SIeqn:latticekineticenergy}, and the relation between bond extensions and degrees of freedom $\vec{q}$: [\cref{SIeqn:nonlinear_extensions,SIeqn:cellbondextension}], we define the Lagrangian:
\begin{align}
    \mathrm{L}(\{q_i,\dot{q}_i\}_i) & := T_{\mathrm{latt}} - E_{\mathrm{latt}} , \label{SIeqn:discretelagrangian}
\end{align}
and solve for dynamics by extremising the associated action, to yield the nonlinear equations of motion (Euler-Lagrange equations):
\begin{align}
    \mathcal{A} \left [\{q_i,\dot{q}_i\}_i\right ] & := \int dt \, \mathrm{L}(\{q_i,\dot{q}_i\}_i), \\
    \frac{\delta \mathcal{A}}{\delta q_i} & = 0 \qquad \qquad \implies \frac{d}{dt} \frac{\partial T_{\mathrm{latt}}}{\partial \dot{q}_i} = - \frac{\partial E_{\mathrm{latt}}}{\partial q_i}.
    \label{SIeqn:nonlin_EOM}
\end{align}

To simulate the experimental conditions, we need to incorporate external driving forces acting on the blocks, say $\vec{F}_{\mathrm{d}}$. Most components of $\vec{F}_{\mathrm{d}}$ are zero-valued except those corresponding to the corner blocks that are clamped by external agents. The experiments also involve dissipation which we assume to be effectively internal, linearly resisting the deformation rate of the ligaments. Using the same symmetry arguments as discussed in \Cref{subsec:hinge_symm_deform}, we find that these internal dissipative forces relate to the bending, stretching and shearing rates with respective dissipation coefficients $\eta_\theta$, $\eta_l$ and $\eta_s$.


\begin{figure*}[h]
\centering
    \includegraphics[width=0.95\textwidth]{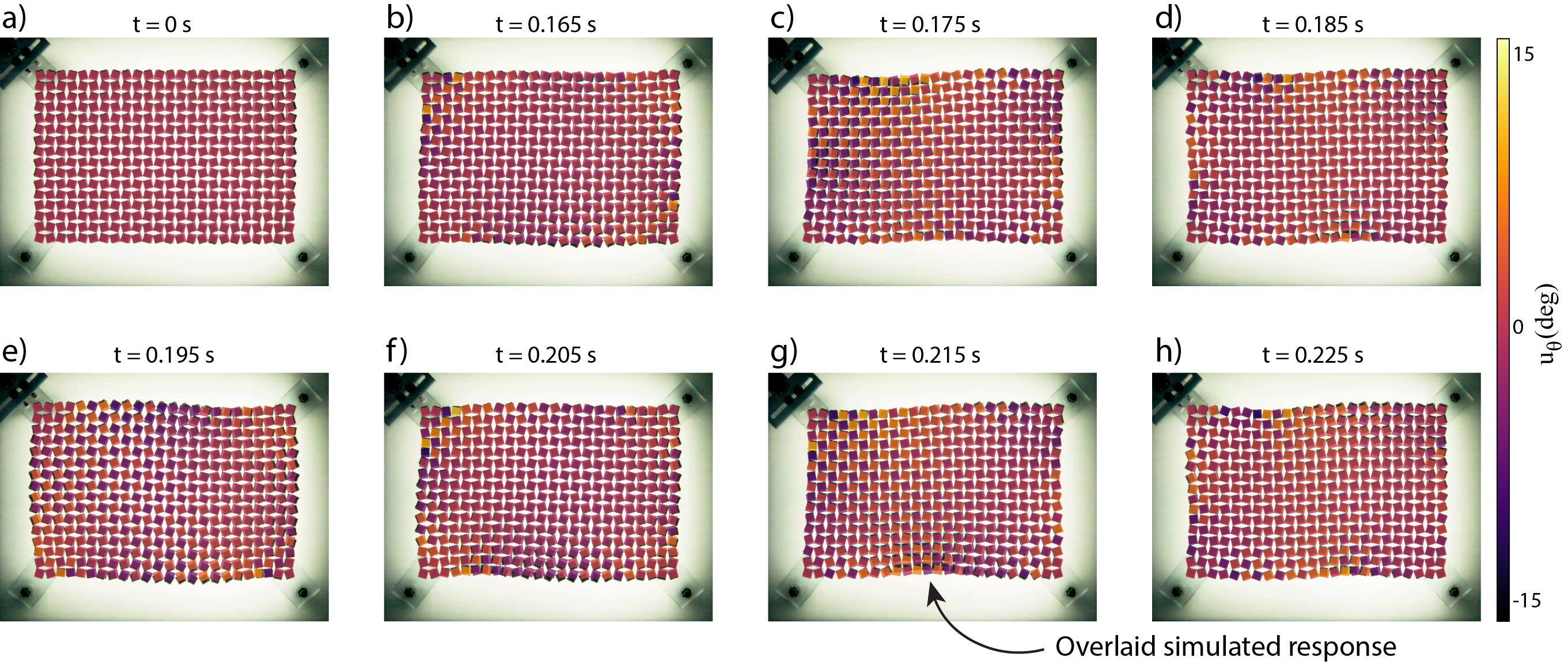}
    \caption{\textbf{Nonlinear Model Validation:} Simulated response overlaid on experimental video a) before excitation is applied and b-h) during excitation at $f=24$ Hz.}
    \label{fig:SI_NL_model_validation}
\end{figure*}

\subsubsection{Nonlinear Discrete Model Validation}
\label{subsubsec:nonlin_discrete_model}

To capture the dynamics of rigid squares connected by thin ligaments at their hinges, we implement the differentiable simulation framework published in \cite{Bordiga2024}. The energy functions in this framework are equivalent to \cref{SIeqn:latticedeformenergy,SIeqn:latticekineticenergy}, using nonlinear formulations for the hinge extensions \cref{SIeqn:nonlinear_extensions}, and including internal dissipation through a Rayleigh dissipation potential,

\begin{equation}
    Q(\vec{q}, \dot{\vec{q}}, t; \vec{\eta}) = -\frac{\partial{C}}{\partial\dot{\vec{q}}}
\end{equation}
where $\vec{q}$ and $\dot{\vec{q}}$ are generalized coordinates, $\vec{\eta}$ contains the damping parameters, and the dissipation potential is written as
\begin{equation}
    C(\vec{q}, \dot{\vec{q}}, t; \vec{\eta}) = \frac{1}{2} \sum_{\dn} \sum_{\alpha=0}^3(\eta_{\theta} (\delta \dot{\theta}_{\alpha}^{\dn})^2 + \eta_l (\delta \dot{l}_{\alpha}^{\dn})^2 + \eta_s (\delta \dot{s}_{\alpha}^{\dn})^2)
\end{equation}
and $\delta \dot{\theta}$, $\delta \dot{l}$, and $\delta \dot{s}$ are the bending, stretching, and shearing hinge deformation rates, respectively, calculated as the time derivatives of the nonlinear hinge extensions \cref{SIeqn:nonlinear_extensions}.



Within this framework, we identify our geometric parameters, including initial angle, $\theta_0$, uniform block spacing, $a$, and hinge length, $l$. The stiffnesses of each ligament are characterized by three stiffness parameters fit to experimental data: an axial stiffness $k_l= 20.59$ N/mm, a shearing stiffness $k_s=0.83$ N/mm, and a rotational stiffness $k_{\theta}=1.68$ N-mm. Furthermore, the damping associated with each ligament is characterized by three damping coefficients associated with the axial velocity $\eta_l = 0.77$ kg/s, shear velocity $\eta_s = 0.77$ kg/s, and rotational velocity $\eta_{\theta}=2.7 \times 10^{-6}$ kg-m$^2$/s between adjacent blocks. \Cref{fig:SI_NL_model_validation}(a-h) highlight the remarkable agreement between experimental video (background) and overlaid simulated response (foreground) of the $\theta_{0}=30^{\circ}$ sample driven at $f=24$ Hz.

\subsubsection{Linear Discrete Model}

In the linear limit, the deformation energy of the lattice is a quadratic form in these generalised coordinates $\vec{q}$ (block degrees of freedom), given by
\begin{align}
    E_{\mathrm{latt}} & \approx \frac{1}{2} \vec{q}^T \mathcal{D} \vec{q}, & \mathcal{D}_{ij} & = \lim_{q_i,q_j \to 0} \frac{\partial^2 E_{\mathrm{latt}}}{\partial q_i \partial q_j}
\end{align}
where $\mathcal{D}$ is the $3N\times 3N$ dynamical matrix, found by substituting the linear approximations for hinge extensions [\cref{SIeqn:cellbondextension}] into the lattice energy expression \cref{SIeqn:latticedeformenergy} and taking its hessian with respect to $\vec{q}$. This approximation neglects geometric nonlinearities arising from hinge kinematics [\cref{SIeqn:nonlinear_extensions}], which become relevant at large amplitudes.

The lattice kinetic energy \cref{SIeqn:latticekineticenergy} is already a quadratic form in the generalised velocities $\dot{\vec{q}} = (\dot{q}_1, \dot{q}_2, \ldots, \dot{q}_{3N})$, with a diagonal generalised mass matrix $\mathcal{M}$ made up of block masses and moment of inertia:
    \begin{flalign}
        && T_{\mathrm{latt}} & = \frac{1}{2} \dot{\vec{q}}^T \mathcal{M} \dot{\vec{q}}, & \mathcal{M}_{ij} & = \frac{\partial^2 T_{\mathrm{latt}}}{\partial \dot{q}_i \partial \dot{q}_j} && \label{SIeqn:lattkineticquad} \\
        \implies && \mathcal{M}_{ij} & = 0 \quad \forall \, i \neq j, & \mathcal{M}_{ii} & = \frac{\partial^2 T_{\mathrm{latt}}}{\partial \dot{q}_i^2} = \begin{cases}
            m & \text{if $q_i$ is a displacement} \\
            J & \text{if $q_i$ is a rotation}
        \end{cases}. &&
    \end{flalign}
where $J = mc^2/6$ is the out-of-plane moment of inertia of a block.

The Euler-Lagrange equations [\cref{SIeqn:nonlin_EOM}] simplify to a linear system describing the small-amplitude dynamics of the lattice:
\begin{align}
    \frac{d}{dt} \frac{\partial T_{\mathrm{latt}}}{\partial \dot{q}_i} & = - \frac{\partial E_{\mathrm{latt}}}{\partial q_i} & \xRightarrow[\text{limit}]{\text{linear}} && \mathcal{M} \ddot{\vec{q}} & = -\mathcal{D} \vec{q} 
    \label{SIeqn:lineardiscreteEOM}.
\end{align}
\textbf{Linear deformation modes} are found as $\vec{q}_n e^{i\omega_n t}$ by numerically solving the eigenvalue problem $-\omega_n^2 \mathcal{M} \vec{q}_n = -\mathcal{D} \vec{q}_n$ for the same parameter values used in nonlinear simulations [\Cref{subsubsec:nonlin_discrete_model}]. Here, $\omega_n$ is the eigenfrequency, and $\vec{q}_n$ is the mode profile for mode index $n \in [3N]$, as reported in main-paper Figure~4.

For damped-driven lattices, such as our experiments, with external drive force $\vec{F}_{\mathrm{d}}$ and a linearised damping matrix $\mathcal{N}$, the linear response is the solution of:
\begin{align}
    \mathcal{M} \ddot{\vec{q}} & = -\mathcal{D} \vec{q} - \mathcal{N} \dot{\vec{q}} + \vec{F}_{\mathrm{d}} .\label{SIeqn:lineardiscretedampeddrivenEOM}
\end{align}
To linearly simulate our \textbf{periodically driven experiments}, we note that the active external force acts on three blocks on the top-left corner at drive frequency $\omega_{\mathrm{d}}$, i.e. $\vec{F}_{\mathrm{d}}(t) := \vec{F}_{\mathrm{d}} e^{i\omega_{\mathrm{d}}t}$. The steady-state linear response oscillates at the same frequency, i.e. $\vec{q}(t) := \vec{q} e^{i \omega_{\mathrm{d}} t}$. Substituting this oscillating response in \cref{SIeqn:lineardiscretedampeddrivenEOM}, we find the following linear system predicting the experimental response:
\begin{align}
    -\omega_{\mathrm{d}}^2 \mathcal{M} \vec{q} & = -\mathcal{D} \vec{q} - i \omega_{\mathrm{d}} \mathcal{N}  \vec{q} + \vec{F}_{\mathrm{d}}.
\end{align}

To linearly simulate the \textbf{relaxing regime in experiments} after active driving stops, we use the experimental snapshot at the moment driving halts as the initial condition $\vec{q}(0), \dot{\vec{q}}(0)$ for solving \cref{SIeqn:lineardiscretedampeddrivenEOM}. Since the external forces are only supplied by the corner-clamp constraints, we can eliminate those degrees of freedom to get a reduced equation for free blocks: $\mathcal{M}_f \ddot{\vec{q}}_f = -\mathcal{D}_f \vec{q}_f - \mathcal{N}_f \dot{\vec{q}}_f$. To find the relaxing response, we first chart the damped modes and complex eigenfrequencies of this reduced linear system. Then, the initial snapshot can be decomposed using the eigenbasis of these damped linear modes, to find the mode coefficients that describe the linear response.

\section{\label{subsec3}Coarse-grained Continuum Model}

We link the linearised discrete dynamics with an effective continuum elasticity theory that provides critical insights into the macroscopic behaviour of such dilational metamaterials. This effective continuum theory is derived by first taking a long-wavelength limit, which involves coarse-graining the six discrete degrees-of-freedom and expanding in gradients, and then restricting to low-frequency regime by eliminating gapped bands to obtain a reduced two-dimensional elasticity theory with an anomalous bulk modulus.

\subsection{Long-Wavelength Approximation}

Since there are $6$ degrees of freedom per unit cell, we adopt the following basis of $6$ continuum degrees of freedom to describe the deformations:
\begin{equation}
\begin{aligned}
    \du(\dr) & := \frac{\du_1(\dr) + \du_2(\dr)}{2} &
    \du_{\Delta}(\dr) & := \frac{\du_1(\dr) - \du_2(\dr)}{2} \\
    \theta_d(\dr) & := \frac{\theta_1(\dr) + \theta_2(\dr)}{2} &
    \theta_r(\dr) & := \frac{\theta_1(\dr) - \theta_2(\dr)}{2}, \label{SIeqn:continuumbasis}
\end{aligned}
\end{equation}
where we are suppressing the explicit time dependence of these fields for a cleaner presentation. The field $\du$ captures the coarse-grained displacements and $\du_{\Delta}$ captures the relative (jagged) displacements between neighbours.
The field $\theta_d$ captures the counter-rotations of neighbours related to bending hinges, whereas $\theta_r$ captures the co-rotations of neighbours accounting for coarse-grained rotation of the larger patch.

For any cell indexed by $\dn$ (see \Cref{fig:SI_lattice_dofs}a), the central square $1$ is located at $\dr_1^{\dn}$ and the square $2$ on its right is located at $\dr_2^{\dn}$, as defined in \cref{SIeqn:refconfig}. The displacements and rotations of these two blocks are expressed in our continuum basis, using \cref{SIeqn:continuumbasis},
\begin{align}
    \begin{aligned}
        \du_1^{\dn} & = \du(\dr_1^{\dn}) + \du_{\Delta}(\dr_1^{\dn}) & \du_2^{\dn} & = \du(\dr_2^{\dn}) - \du_{\Delta}(\dr_2^{\dn}) \\
        \theta_1^{\dn} & = \theta_{d}(\dr_1^{\dn}) + \theta_{r}(\dr_1^{\dn}) & \theta_2^{\dn} & =  \theta_{d}(\dr_2^{\dn}) - \theta_{r}(\dr_2^{\dn}).
    \end{aligned} \label{SIeqn:continuumdofs}
\end{align}
We use this to find the hinge-deformations of cell $\dn$ [\cref{SIeqn:cellbondextension}] as a function of the continuum fields. The extensions of the hinge at corner $\alpha$ in cell $\dn$ (see \Cref{fig:SI_lattice_dofs}a) are expressed as:
    \begin{subequations}
    \begin{align}
        \delta \theta_{\alpha}^{\dn} & =  \theta_{d}(\dr_2^{\dn_{\alpha}}) + \theta_{d}(\dr_1^{\dn}) + \theta_{r}(\dr_1^{\dn}) - \theta_{r}(\dr_2^{\dn_{\alpha}})  \\
        \delta l_{\alpha}^{\dn} & = \Big [ \du(\dr^{\dn_\alpha}_{2}) - \du(\dr^{\dn}_{1}) - \du_{\Delta}(\dr^{\dn_\alpha}_{2})  - \du_{\Delta}(\dr^{\dn}_{1}) \Big ] \cdot \hat{l}_{\alpha} +\frac{a-l}{2} \tan{\theta_0} \Big [ \theta_{d}(\dr_1^{\dn}) + \theta_{d}(\dr_2^{\dn_\alpha}) + \theta_{r}(\dr_1^{\dn})  - \theta_{r}(\dr_2^{\dn_\alpha}) \Big ] \\
        \delta s_{\alpha}^{\dn} & = \hat{z} \cdot \Big ( \hat{l}_{\alpha} \times \left [ \du(\dr^{\dn_\alpha}_{2}) - \du(\dr^{\dn}_{1}) - \du_{\Delta}(\dr^{\dn_\alpha}_{2}) - \du_{\Delta}(\dr^{\dn}_{1}) \Big ] \right ) + \frac{a}{2} \Big [ \theta_{d}(\dr_2^{\dn_\alpha}) - \theta_{d}(\dr_1^{\dn}) - \theta_{r}(\dr_2^{\dn_\alpha}) - \theta_{r}(\dr_1^{\dn}) \Big ],
    \end{align}
    \end{subequations}
where $\dn_{\alpha}$ denotes cell index of neighbouring square connected at corner $\alpha$ of cell $\dn$, as described in \Cref{tab:cellneighbourindex}.

With these equations, we can approximate the energetics of the system in terms of the continuum fields under the assumption of a slowly varying deformation. We briefly note the Taylor expansion of these continuum fields about the position of a reference square (taken to be $\dr^{\dn}_1$), which are used in the homogenisation of the unit cell energies,
\begin{align}
    \text{for } f & \in \{ u_x, u_y, u_{\Delta x}, u_{\Delta y}, \theta_{d}, \theta_{r} \} \nonumber \\
    f(\dr^{\dn_\alpha}_{2}) & = f(\dr^{\dn}_{1} + a\hat{l}_{\alpha}) 
    \approx f(\dr^{\dn}_{1}) + a\hat{l}_{\alpha,i} \partial_i f \mid_{\dr^{\dn}_{1}} + \frac{a^2}{2} \hat{l}_{\alpha,i} \hat{l}_{\alpha,j} \partial_i \partial_j f \mid_{\dr^{\dn}_{1}}  + \mathcal{O}\left(\partial^3\right), \label{SIeqn:taylorseries}
\end{align}
where the higher-order gradients can be ignored in the long-wavelength limit: $\lambda \sim \frac{\|f\|}{\|\nabla f \|} \gg a$. 

\subsubsection{Deformation Energy Density}
Using the above equations, we express the bending, stretching and shearing energies of cell $\dn$ from \cref{SIeqn:celldeformenergy} using the continuum fields, Tayor-expanded about $\dr_1^{\dn}$,
\begin{subequations} \label{SIeqn:cellcontinuumenergy}
    \begin{align}
        \text{bend energy:} && \frac{k_{\theta}}{2} \sum_{\alpha=0}^3 (\delta \theta_{\alpha}^{\dn})^2 & =k_{\theta} \bigg [ 8 \theta_{d}^2 + a^2 \left ( \nabla \left ( \theta_{d} - \theta_{r} \right ) \right )^2 + 2a^2 \theta_{d} \nabla^2 \left ( \theta_{d} - \theta_{r} \right ) + \mathcal{O}\left(\partial^3\right) \bigg ]_{\dr=\dr_1^{\dn}} ,\label{SIeqn:cellbendcontinuumenergy} \\
        \text{stretch energy:} && \frac{k_l}{2}\sum_{\alpha=0}^3 (\delta l_{\alpha}^{\dn})^2
        & = \frac{k_l}{2} \bigg [ 8 \du_{\Delta}^2 + 4(a-l)^2 \tan^2\theta_0 \, \theta_{d}^2 + a^2 \left (
        \nabla \cdot (\du - \du_{\Delta}) \right )^2 +  4 a (a-l) \tan\theta_0 \, \theta_{d} \nabla \cdot \left ( \du - \du_{\Delta} \right) \nonumber \\
        && & + a^2 \left (\partial_x(u_x - u_{\Delta,x}) - 
        \partial_y(u_y - u_{\Delta,y}) \right )^2 + 4 a^2 \du_{\Delta} \cdot \left ( \partial_x^2 (u_{\Delta,x} - u_x ) \hat{x} +
         \partial_y^2 (u_{\Delta,y} - u_y) \hat{y} \right ) \nonumber \\
        && & + 4 a (a-l) \tan\theta_0 \, \du_{\Delta} \cdot \nabla \left ( \theta_{r} - \theta_{d} \right ) + \frac{a^2(a-l)^2 \tan^2\theta_0}{2}
        \nabla \left ( \theta_{d} - \theta_{r} \right ) \cdot \nabla \left ( \theta_{d} - \theta_{r} \right )  \nonumber \\
        && & + a^2(a- l)^2 \tan^2\theta_0  \, \theta_{d} \nabla^2 \left( \theta_{d} - \theta_{r}\right ) + \mathcal{O}\left(\partial^3\right) \bigg ]_{\dr=\dr_1^{\dn}} \label{SIeqn:cellstretchcontinuumenergy} , \\
        \allowdisplaybreaks
        \text{shear energy:} && \frac{k_s}{2} \sum_{\alpha=0}^3 (\delta s_{\alpha}^{\dn})^2 & = \frac{k_s}{2} \bigg [ 8 \du_{\Delta}^2 + 4a^2  \theta_{r}^2 + a^2 \left ( \nabla \times (\du - \du_{\Delta}) \right )^2 +  4a^2 \theta_{r} \left ( \nabla \times (\du - \du_{\Delta}) \right )_z \nonumber   \\
        && & + a^2 \left ( \partial_y (u_x - u_{\Delta,x}) + \partial_x (u_y - u_{\Delta,y}) \right )^2 + 4a^2 \du_{\Delta} \cdot \left ( \partial_y^2 (u_{\Delta,x} - u_x ) \hat{x} +
         \partial_x^2 (u_{\Delta,y} - u_y) \hat{y} \right ) \nonumber \\
        && & - 4a^2 (\du_{\Delta} \times \nabla \left ( \theta_{d}  - \theta_{r} \right ))_z + \frac{a^4}{2}
        \nabla \left ( \theta_{d} - \theta_{r} \right ) \cdot \nabla \left ( \theta_{d} - \theta_{r} \right ) \nonumber \\
        && & - a^4 \theta_{r} \nabla^2 \left (\theta_{d} - \theta_{r} \right ) + \mathcal{O}\left(\partial^3\right) \bigg ]_{\dr=\dr_1^{\dn}}  .\label{SIeqn:cellshearcontinuumenergy}
    \end{align}
\end{subequations}
The continuum deformation energy density interpolates the cell deformation energy per unit area,
\begin{align}
    E(\dr_1^{\dn}) & = \frac{E_{\mathrm{cell}}[\dn]}{A_{\mathrm{cell}}} = \frac{1}{ 2a^2 } \left ( \frac{k_{\theta}}{2} \sum_{\alpha} (\delta \theta_{\alpha}^{\dn})^2 + \frac{k_l}{2} \sum_{\alpha} (\delta l_{\alpha}^{\dn})^2 + \frac{k_s}{2} \sum_{\alpha} (\delta s_{\alpha}^{\dn})^2 \right ).
\end{align}
Thus, we find the continuum deformation energy density $E(\dr)$ by adding up the energy expressions of \cref{SIeqn:cellcontinuumenergy} for cell located at $\dr$ as
\begin{align}
    \implies E(\dr) & = 2\frac{k_l+k_s}{a^2} \du_{\Delta}^2 + \frac{ 4k_{\theta}+ k_l(a-l)^2 \tan^2\theta_0 }{a^2} \left ( \theta_{d} + \frac{k_la(a-l)\tan\theta_0}{8k_{\theta}+2k_l(a-l)^2\tan^2\theta_0} \nabla \cdot \left ( \du - \du_{\Delta} \right)  \right )^2 \nonumber \\
    & + \frac{k_{\theta}k_l }{4k_{\theta} + k_l (a-l)^2 \tan^2\theta_0} \left ( \nabla \cdot (\du - \du_{\Delta}) \right )^2 + \frac{k_s}{4} \left (2\theta_{r} + \left ( \nabla \times (\du - \du_{\Delta}) \right )_z  \right )^2 \nonumber \\
    & + \frac{k_l}{4} \left (\partial_x(u_x - u_{\Delta,x}) - 
    \partial_y(u_y - u_{\Delta,y}) \right )^2 + \frac{k_s}{4}\left (\partial_y(u_x - u_{\Delta,x}) + 
    \partial_x(u_y - u_{\Delta,y}) \right )^2 \nonumber \\
    & + \left ( \frac{k_{\theta}}{2} + \frac{(a-l)^2 k_l \tan^2\theta_0 + a^2 k_s}{8} \right ) \left ( \nabla \left ( \theta_{d} - \theta_{r} \right ) \right )^2 + \left ( k_{\theta} \theta_{d} + \frac{(a-l)^2}{4}k_l \tan^2\theta_0\, \theta_{d} - \frac{a^2}{4}k_s \theta_{r} \right ) \nabla^2 \left ( \theta_{d} - \theta_{r} \right ) \nonumber \\
    & + k_l \du_{\Delta} \cdot \left ( \partial_x^2 (u_{\Delta,x} - u_x ) \hat{x} +
    \partial_y^2 (u_{\Delta,y} - u_y) \hat{y} \right ) + k_s \du_{\Delta} \cdot \left ( \partial_y^2 (u_{\Delta,x} - u_x )\hat{x} +
    \partial_x^2 (u_{\Delta,y} - u_y) \hat{y} \right ) \nonumber\\
    & - k_l \frac{a-l}{a} \tan\theta_0 \, \du_{\Delta} \cdot \nabla \left ( \theta_{d} - \theta_{r} \right ) - k_s \tan\theta_0 \, \left (\du_{\Delta} \times \nabla \left ( \theta_{d} - \theta_{r} \right ) \right )_z + \mathcal{O}\left(\partial^3\right). \label{SIeqn:deformenergydensity}
\end{align}

\subsubsection{Kinetic Energy Density}
Similarly, we can express the cell kinetic energy of \cref{SIeqn:cellkineticenergy} in terms of the continuum fields [\cref{SIeqn:continuumdofs}],
    \begin{align}
        T_{\mathrm{cell}}[\dn] & = \frac{m}{2} \left ( \dot{\du}(\dr^{\dn}_1) + \dot{\du}_{\Delta}(\dr^{\dn}_1)  \right)^2 + \frac{m}{8} \sum_{\alpha} \left ( \dot{\du}(\dr^{\dn_{\alpha}}_2) - \dot{\du}_{\Delta}(\dr^{\dn_{\alpha}}_2) \right)^2  + \frac{J}{2}  \left ( \dot{\theta}_{r}(\dr^{\dn}_1) + \dot{\theta}_{d}(\dr^{\dn}_1) \right)^2 + \frac{J}{8} \sum_{\alpha} \left ( \dot{\theta}_{r}(\dr^{\dn_{\alpha}}_2) - \dot{\theta}_{d}(\dr^{\dn_{\alpha}}_2)\right)^2.
    \end{align}
Substituting Taylor-expansions from \cref{SIeqn:taylorseries} into the cell kinetic energy, we find the long-wavelength approximation 
    \begin{align}
        T_{\mathrm{cell}}[\dn] & = m \left ( \dot{\du}^2 + \dot{\du}_{\Delta}^2  + \mathcal{O}\left(\partial^2\right) \right ) \Big \vert_{\dr=\dr^{\dn}_1} +  J \left ( \dot{\theta}_{r}^2 + \dot{\theta}_{d}^2 + \mathcal{O}\left(\partial^2\right)  \right )\Big \vert_{\dr=\dr^{\dn}_1}. 
    \end{align}
    Finally, we find the continuum kinetic energy density $T(\dr)$ that interpolates the cell kinetic energy per unit area:
    \begin{align}
        T(\dr_1^{\dn}) & = \frac{T_{\mathrm{cell}}[\dn] }{A_{\mathrm{cell}}} = \frac{1}{2a^2} \Bigg ( m \left ( \dot{\du}^2 + \dot{\du}_{\Delta}^2  \right) +  J \left ( \dot{\theta}_{r}^2 + \dot{\theta}_{d}^2 \right) +  \mathcal{O}\left(\partial^2\right) \Bigg )_{\dr=\dr_1^{\dn}} \nonumber \\
        \implies T(\dr) & = \frac{m}{2a^2} \left ( \dot{\du}^2 + \dot{\du}_{\Delta}^2  \right) +  \frac{J}{2a^2} \left ( \dot{\theta}_{r}^2 + \dot{\theta}_{d}^2 \right) +  \mathcal{O}\left(\partial^2\right). \label{SIeqn:kineticenergydensity}
    \end{align}

\subsection{Low-frequency Approximation --- RS Elasticity Theory}
\label{subsec:elasticitytheory}
In the low-frequency long-wavelength limit, we expect the dynamics to resolve onto a two-dimensional manifold of acoustic dispersion bands that extend from the zero modes of uniform translation represented as $\du=\mathrm{const.},\, \du_{\Delta}=\mathbf{0},\,  \theta_d=\theta_r=0$ in our continuum basis.
These acoustic bands are described by perturbations about these zero modes that involve slowly-varying coarse displacements $\du$ generating forces proportional to their gradients (linear strains), while the deviatoric displacements $\du_{\Delta}$ remain negligible since they directly stretch and shear the hinges (see \cref{SIeqn:deformenergydensity}).
We will show how the rotational degrees of freedom, $\theta_{d}$ and $\theta_{r}$, follow approximately the gradients of coarse displacements in these low-frequency acoustic modes. Ultimately, this insight will help us choose a modified basis that clearly separates into acoustic and other gapped bands.

Consider a small in-plane macroscopic deformation, applied to an RS lattice, described by a linear strain tensor ($\boldsymbol{\epsilon}$) and rotation tensor ($\boldsymbol{\omega}$), quasi-static in time and slowly-varying in space (uniform over multiple unit cells). Then the coarse displacement is described, up to a uniform translation, as:
\begin{align}
    \du(\dr) & = \left (\boldsymbol{\epsilon} + \boldsymbol{\omega} \right ) \cdot\dr = \frac{1}{2}  \begin{bmatrix}
        d + s_1 & s_2 - r \\ s_2 + r & d - s_1
    \end{bmatrix} \dr
\end{align}
where dilation $d$ and shears $s_1, s_2$ are the orthogonal linear strains, and $r$ is the local rotation, defined by
\begin{gather}\label{SIeqn:linearstrain&rotn}
    \begin{aligned}
            d & :=  \partial_x u_x + \partial_y u_y & \qquad \qquad \qquad r & := \partial_x u_y - \partial_y u_x  \\
            s_1 & := \partial_x u_x - \partial_y u_y & \qquad \qquad \qquad s_2 & := \partial_x u_y + \partial_y u_x
    \end{aligned}
\end{gather}
relating them to the strain and rotation tensor,
\begin{flalign}
        && \boldsymbol{\epsilon} & = \frac{1}{2}  \begin{bmatrix}
            d + s_1 & s_2 \\ s_2 & d - s_1
        \end{bmatrix} && \text{and}
        & \boldsymbol{\omega} & = \frac{r}{2}  \begin{bmatrix}
            0 & -1 \\ 1 & 0
        \end{bmatrix}. &&
\end{flalign}
Here, the angle of rotation corresponding to $\boldsymbol{\omega}$ is half the rotation component $r$.

In the quasi-static limit, the other degrees of freedom, $\du_{\Delta},\theta_d$ and $\theta_r$, must relax to achieve the minimum deformation energy possible for the prescribed strains. This energy-minimising relaxation will indicate how the squares rotate and displace in relation to the coarse displacement field $\du$ at low frequencies.
Since the applied deformation is uniform over several unit cells, we can assume each unit cell relaxes in almost the same way as its neighbours to achieve the lowest energy configuration. As such the gradients of $\du_{\Delta},\theta_d$ and $\theta_r$ can be ignored in the deformation energy calculation in comparison to their direct energy costs in \cref{SIeqn:deformenergydensity}. This gives the simplified energy expression:
\begin{align}
    E_{\mathrm{latt}} \bigg \vert_{\du=\left (\boldsymbol{\epsilon} + \boldsymbol{\omega} \right ) \cdot\dr} & = \int d^2\dr \, E(\dr) \bigg \vert_{\du=\left (\boldsymbol{\epsilon} + \boldsymbol{\omega} \right ) \cdot\dr} \nonumber \\
    & \approx \int d^2 \dr \, \bigg [ \frac{k_{\theta}k_l }{4k_{\theta} + k_l (a-l)^2 \tan^2\theta_0} d^2 + \frac{k_l}{4} s_1^2 + \frac{k_s}{4} s_2^2 + 2\frac{k_l+k_s}{a^2} \du_{\Delta}^2   \nonumber \\
    & + \frac{ 4k_{\theta}+ k_l(a-l)^2 \tan^2\theta_0 }{a^2} \left ( \theta_{d} + \frac{k_la(a-l)\tan\theta_0}{8k_{\theta}+2k_l(a-l)^2\tan^2\theta_0} d  \right )^2 + \frac{k_s}{4} \left (2\theta_{r} - r  \right )^2 \bigg ].
\end{align}
This energy is minimum when the relative displacement field $\du_{\Delta}$ is negligibly excited, and the counter-rotation and co-rotation fields respectively follow the coarse-grained dilation and rotation:
\begin{subequations}\label{SIeqn:relaxed_condns}
\begin{align}
    \du_{\Delta} & = \mathbf{0} \label{SIeqn:zerojaggedness} \\
    \theta_{d} & = - \frac{k_la(a-l)\tan\theta_0}{8k_{\theta}+2k_l(a-l)^2\tan^2\theta_0} d \label{SIeqn:goodbending} \\
    \theta_{r} & = \frac{r}{2}. \label{SIeqn:goodrotating}
\end{align}
\end{subequations}
The coefficient before dilation $d$ in \cref{SIeqn:goodbending} accounts for the inter-play between hinge-bending and hinge-stretching movements of adjacent blocks to accommodate isotropic expansions/compressions.
For lattice geometry with small initial angle $\theta_0$ (like $5^{\circ}$) dilation decouples from the low-energy hinge-bending rotations $\theta_d$, and instead causes the high-energy stretching of hinges.
But for large enough initial angle, tiny hinges ($a\ll l$) and negligible bending stiffness $k_{\theta}$, the strong coupling allows counter rotations $\theta_d$ to accommodate almost all of the dilation, capturing the ideal RS mechanism:
\begin{align}
    \theta_d & \approx -\frac{d}{2\tan{\theta_0}} + \mathcal{O}\left ( \frac{k_{\theta}}{a^2 k_l \tan^2{\theta_0}}\right ) + \mathcal{O}\left ( \frac{l}{a}\right ).
\end{align}
    
Hence, we change our basis of continuum degrees of freedom to differentiate between the manifold of deformations that minimise the potential energy as shown above and the deviations thereof. Let the deviations of $\theta_d$ and $\theta_r$ from \cref{SIeqn:relaxed_condns} be denoted by the deviatoric fields:
    \begin{subequations}
    \begin{align}
        \Delta \theta_{d}(\dr) & := \theta_{d}(\dr) + \frac{k_la(a-l)\tan\theta_0}{8k_{\theta}+2k_l(a-l)^2\tan^2\theta_0} d(\dr) \\
        \Delta \theta_{r}(\dr) & := \theta_{r}(\dr) -  \frac{r(\dr)}{2}.
    \end{align}
    \end{subequations}
Then, the new basis is represented as $\{\du, \du_{\Delta},\Delta\theta_{d},\Delta\theta_{r}\}$.
The deformation energy is orthogonal in this new basis up to higher order gradient terms:
    \begin{align}
        E(\dr) & = \frac{k_{\theta}k_l }{4k_{\theta} + k_l (a-l)^2 \tan^2\theta_0} d^2 + \frac{k_l}{4} s_1^2  + \frac{k_s}{4} s_2^2 +  \mathcal{O}\left(\partial^3 \du\right) \nonumber \\
        & + 2\frac{k_l+k_s}{a^2} \du_{\Delta}^2 + \frac{ 4k_{\theta}+ k_l(a-l)^2 \tan^2\theta_0 }{a^2} \Delta\theta_{d}^2 + k_s \Delta\theta_{r}^2 + \mathcal{O}\left(\partial( \Delta f)\right), \label{SIeqn:lowbandenergy}
    \end{align}
where $\Delta f$ represents each of the deviatoric fields $\{\du_{\Delta},\Delta\theta_{d},\Delta\theta_{r}\}$, and the linear strains $d,s_1,s_2$ were defined earlier in \cref{SIeqn:linearstrain&rotn}. Similarly, the kinetic energy density also turns out to be orthogonal in this basis:
\begin{align}
    T(\dr) & =  \frac{m}{2a^2} \dot{\du}^2   + \frac{m}{2a^2} \dot{\du}_{\Delta}^2  + \frac{J}{2a^2} \Delta \dot{\theta}_{r}^2 + \frac{J}{2a^2} \Delta \dot{\theta}_{d}^2  + \mathcal{O}\left(\partial f\right),\label{SIeqn:lowbandkinetic}
\end{align}   
where $f$ represents any field in the new basis $\{\du, \du_{\Delta},\Delta\theta_{d},\Delta\theta_{r}\}$.

Thus, the new basis shows how the dynamics break up into orthogonal sectors of $\du$, $ \du_{\Delta}$, $\Delta\theta_{d}$ and $\Delta\theta_{r}$, with distinct dispersion bands corresponding each in the long-wavelength limit.
The bands corresponding excitations in $\du$ are gapless (acoustic) since their dynamics are mediated by gradients, whereas the other four bands are gapped away.
Hence, the low-frequency dynamics of RS metamaterial is described by a $2$D elasticity theory concerned only with the gradients of the coarse displacement field, with Lagrangian density
\begin{align}
    \mathcal{L}(\partial_{\mu} \du) & = T - E = \frac{m}{2a^2} \dot{\du}^2 - \left ( \frac{k_{\theta}k_l }{4k_{\theta} + k_l (a-l)^2 \tan^2\theta_0} d^2 + \frac{k_l}{4} s_1^2  + \frac{k_s}{4} s_2^2 \right ),
\end{align}
where neighbouring squares move together [\cref{SIeqn:zerojaggedness}], counter-rotate to accommodate the local dilation [\cref{SIeqn:goodbending}] and co-rotate to accommodate coarse-grained patch rotation [\cref{SIeqn:goodrotating}]. 

This is an anisotropic elasticity theory with the relevant effective elastic moduli and mass density given by:
\begin{subequations} \label{SIeqn:moduli}
\begin{flalign}
        \text{mass density : } && \rho & = \frac{m}{a^2} && \label{SIeqn:massdensity} \\
        \text{bulk modulus : } && B & = \frac{2k_{\theta}k_l }{4k_{\theta} + k_l (a-l)^2 \tan^2\theta_0} \approx \frac{2k_{\theta}}{a^2 \tan^2{\theta_0}} + \mathcal{O}\left ( \frac{k_{\theta}}{a k_l}\right ) + \mathcal{O}\left ( \frac{l}{a}\right )  \\
        \text{shear moduli : } && G_1 & = \frac{k_l}{2}, 
        \qquad \qquad \qquad \qquad G_2 = \frac{k_s}{2} \label{SIeqn:elastic_moduli}
\end{flalign}
\end{subequations}
where we have approximated the bulk modulus for tiny hinges $l\ll a$ that bend easily $k_{\theta}\ll a^2 k_l$. 

\subsubsection{First-Order Lattice Effects}
A possible consequence of highly dilational deformations, witnessed in our discrete model simulations of $30^{\circ}$ design, is that the energy cost of dilation gradients ($\partial_i d$) becomes quite significant compared to the direct energy cost of dilation. We can infer this analytically by including up to order-4 gradients of $\du$ in its Taylor series \cref{SIeqn:taylorseries} for continuum energy calculation. The resulting deformation energy function for the acoustic bands has an additional term for dilation gradients $\nabla d$:
\begin{align}
    E(\dr) & = \frac{B}{2} d^2 + \frac{G_1}{2} s_1^2  + \frac{G_2}{2} s_2^2 + \frac{M}{2} a^2(\nabla d)^2 +  \mathcal{O}\left(\partial_j d \partial_k s_i\right), \\
    M & = \frac{k_{\theta}}{4a^2} + \frac{k_s}{16} + \frac{k_l}{16} \left (\tan^2\theta_0 + \frac{16k_{\theta}^2 + a^2k_l\tan^2\theta_0 (4 k_{\theta} + a^2k_s)}{(4 k_{\theta} + a^2k_l\tan^2 \theta_0)^2} \right ) + \mathcal{O}\left (\frac{l}{a} k_l \right ), \label{SIeqn:dilngrad_modulus}
\end{align}
where, for brevity, the coefficient of the dilation-gradient term, modulus $M$, has been approximated for tiny hinge-size $l\ll a$ and large stretching stiffness $k_l \gg k_s$.

In the experimental RS metamaterial samples, with their empirical hinge stiffnesses given in \Cref{subsubsec:nonlin_discrete_model}, the stretching stiffness is much larger than others: $k_{\theta}/a^2 \ll k_s \ll k_l$. To compare the relative energy trade-off between the dilation-gradient and dilation terms for such hard-to-stretch hinges, consider a lattice with $N$ blocks and a characteristic length $L=\sqrt{N}a$. Then, for a deformation of characteristic wavenumber $n$ (wavelength $\lambda \sim L/n$), the relative gradient of dilation scales as $\nabla d/ d \sim n/L$. Consequently, the ratio of energy costs from the dilation-gradient and dilation terms scale as:
\begin{align}
    \frac{E_{\nabla d}}{E_d} & = \frac{Ma^2 (\nabla d)^2}{B d^2} \sim \frac{Ma^2 n^2}{BL^2} \sim \frac{n^2}{N} \frac{a^2 k_l}{k_{\theta}} \tan^2\theta_0,
\end{align}
where $M\sim k_l$, from \cref{SIeqn:dilngrad_modulus}, for a large $k_l$.

Thus, we find that the dilation-gradients can only be ignored for deformations with $n^2 \ll N k_{\theta}/(a^2 k_l \tan^2\theta_0)$. For the experimental sample of $\theta_0=30^{\circ}$ design, with $N=384$ and $k_{\theta}/a^2 k_l \approx 0.0008$, we find this threshold to be $N k_{\theta}/(a^2 k_l \tan^2\theta_0) \approx 1 $. Thus, at this experiment size, we expect even the lowest deformation modes to be affected by the cost of dilation-gradients.

In contrast, the relative energy cost of shear-gradients vs shears has a straightforward trade-off of $n^2/N$ which can be ignored for wavelengths much longer than cell size: $\lambda \gg L/\sqrt{N} \approx a$. Similarly, the velocity-gradient terms in the kinetic energy density [\cref{SIeqn:kineticenergydensity}] that arise from block rotations, corresponding $\dot{\theta}_d^2 \sim \dot{d}^2$ and $\dot{\theta}_r^2 \sim \dot{r}^2$, are negligible for wavelengths spanning more than a couple unit cells ($\lambda\gg a$).


Hence, the low-frequency linear coarse-grained elasticity theory for the RS metamaterial in our experiments is aptly described by the energy and Lagrangian densities:
\begin{align}
    E(\partial_i \du, \partial_i \partial_j \du) & = \frac{B}{2} d^2 + \frac{G_1}{2} s_1^2  + \frac{G_2}{2} s_2^2 + \frac{M}{2N} (L\nabla d)^2 \label{SIeqn:RSdeformenergy} \\
    T(\dot{\du}) & = \frac{\rho}{2} \dot{\du}^2 \label{SIeqn:RSkineticenergy} \\
    \mathcal{L}(\dot{\du}, \partial_i \du, \partial_i \partial_j \du) & = T(\dot{\du}) - E(\partial_i \du, \partial_i \partial_j \du) \label{SIeqn:RSlagrangian}
\end{align}
where the mass density and elastic moduli are given in \cref{SIeqn:moduli} and \cref{SIeqn:dilngrad_modulus}, and lattice area $L^2 =N a^2$.

\section{Conformal $2$D Elastodynamics}
\label{sec:conformalmodel}
In this section, we define the complex mapping used to represent the linear conformal elasticity theory that describes deformations of RS metamaterial. We show how the low-frequency near-conformal modes and high-frequency non-conformal bulk modes are found in the dispersion. We define a measure for calculating the non-conformal deviation of a deformation from exact conformal nature, which is used to verify the conformal nature of simulated and experimental RS deformations in the Main Paper (Fig.~4c,d,e and Fig.~5a). Finally, we derive the conservation laws corresponding to the continuous conformal symmetry stated in the Main Paper, and discuss the deviations that arise due to explicit symmetry breaking.

\subsection{Complex Mapping}
The real material coordinates $\langle x,y \rangle$ are mapped to the complex scalar $z=x+iy$.
The conjugate $\bar{z} = x-iy$ is considered a complex coordinate independent from $z$, such that we have a linear map from $\langle x,y \rangle$ to $\langle z,\zb \rangle$. Using the imposed independence, $\partial_z \bar{z} = \partial_{\bar{z}} z = 0$, in addition to $\partial_z z = \partial_{\bar{z}} \bar{z} = 1$, we find the complex partial (Wirtinger) derivatives:
\begin{align}
    \partial_z & = \frac{\partial_x - i\partial_y}{2} & \partial_{\bar{z}} & = \frac{\partial_x + i\partial_y}{2}.
    \label{SIeqn:complexpartials}
\end{align}

Similar to material coordinates, the real in-plane displacements are mapped to a complex scalar field $u(z,\bar{z},t) = u_x(x,y,t) + iu_y(x,y,t)$ and its complex conjugate $\bar{u} = u^* = u_x(x,y,t) - iu_y(x,y,t) $. The gradient components of the displacement field, linear strains and rotation \cref{SIeqn:linearstrain&rotn}, are expressed as complex gradients of complex scalar field $u(z,\zb)$, using \cref{SIeqn:complexpartials}, 
\begin{align}
    2\partial_z u & = \partial_x u_x + \partial_y u_y + i(\partial_x u_y - \partial_y u_x) = d + ir \\
    2\partial_{\bar{z}} u & = \partial_x u_x - \partial_y u_y + i(\partial_x u_y + \partial_y u_x) = s_1 + is_2.
    \label{SIeqn:complexstrains}
\end{align}

This distinction between holomorphic gradients (dilations and rotations) and anti-holomorphic gradients (shears) naturally points to two contrasting special cases for deformations: purely shearing deformations $z \mapsto z + u(\zb) $ that preserve local area and orientation but not local shape (angles), and purely dilational deformations $z \mapsto z + u(z)$ that only preserve local shape but not area or orientation.
The latter class corresponds to conformal mappings of the system, $f(z) = z + u(z)$, that cost deformation energy at the order of their first derivative through bulk modulus $B$ and the second derivative through the dilation-gradient modulus $M$.
This can be readily inferred from the continuum deformation energy density of RS metamaterial [\cref{SIeqn:RSdeformenergy}] represented in complex notation:
\begin{align}
    E(\partial_i u, \partial_i \bar{u}, \partial_i \partial_j u,  \partial_i \partial_j \bar{u}) & = \frac{G_1}{2} (\partial_{\bar{z}} u + \partial_{z} \bar{u})^2  - \frac{G_2}{2} (\partial_z \bar{u} - \partial_{\bar{z}} u)^2 + \frac{B}{2} (\partial_z u + \partial_{\bar{z}} \bar{u})^2 \nonumber \\
    & \quad + \frac{8M}{N} L^2 \partial_z (\partial_z u + \partial_{\bar{z}} \bar{u}) \partial_{\bar{z}} (\partial_z u + \partial_{\bar{z}} \bar{u}) \label{SIeqn:complexpotential} \\
    & = E_{\mathrm{conf}}(\partial_{\zb} u, \partial_z \bar{u}) + \mathcal{O}\left(B\right) + \mathcal{O}\left(M/N\right) \label{eqn:conformalpotential}
\end{align}
where $E_{\mathrm{conf}}$ is the energy cost of shear strains. The kinetic energy density takes the simple complex form:
\begin{align}
    T(\dot{u},\dot{\ub}) & = \frac{\rho}{2} \dot{u} \dot{\ub} .\label{SIeqn:complexkinetic}
\end{align}

\Cref{eqn:conformalpotential} shows that, for an ideal-dilational material with $B \to 0$ and $1/N \to 0$, conformal deformations cost zero energy giving an energy-free degenerate manifold of displacement fields with zero shear. The Lagrangian density, in complex notation, can similarly be expressed as an ideal-limit conformal Lagrangian density $\mathcal{L}_{\mathrm{conf}}$ with two added perturbative terms:
\begin{flalign}
    && \mathcal{L} & =  T(\dot{u}, \dot{\bar{u}}) - E(\partial_i u, \partial_i \bar{u}, \partial_i \partial_j u,  \partial_i \partial_j \bar{u}) = \mathcal{L}_{\mathrm{conf}} + \mathcal{O}\left(B\right) +\mathcal{O}\left(M/N\right) && \label{SIeqn:perturbL} \\
    \text{where} && \mathcal{L}_{\mathrm{conf}} & =  T(\dot{u}, \dot{\bar{u}}) - E_{\mathrm{conf}}(\partial_{\zb} u, \partial_z \bar{u}). \label{SIeqn:confL}
\end{flalign}

The equations of motion are found by extremising the associated action $\mathcal{A}$ with respect to $u$ and $\bar{u}$:
\begin{flalign}
        && \mathcal{A}[u,\bar{u}] & = \int \mathcal{L} \,dzd\bar{z} dt = \int  \left ( \mathcal{L}_{\mathrm{conf}} + \mathcal{O}\left(B\right) +\mathcal{O}\left(M/N\right) \right )  \,dzd\bar{z} dt &&  \\
        \text{extremising action:} && \frac{\delta \mathcal{A}}{\delta u} & = 0 \qquad \text{and} \qquad \frac{\delta \mathcal{A}}{\delta \bar{u}} = 0 \\
        \implies &&  \partial_{\mu} \left ( \frac{\partial \mathcal{L}}{\partial (\partial_{\mu}u)} \right ) & = 0 \qquad \text{and} \qquad  \partial_{\mu} \left ( \frac{\partial \mathcal{L}}{\partial (\partial_{\mu}\ub)} \right ) = 0 .\label{SIeqn:eulerlagrangecomplex}
\end{flalign}
For real displacement fields $u_x$ and $u_y$, $u$ and $\bar{u}$ are complex conjugates of each other, making the two equations of motion [\cref{SIeqn:eulerlagrangecomplex}] complex conjugates of each other, and hence redundant. So without any loss of information, we pick one of them to describe the dynamics:
\begin{align}
    \partial_{\mu} \left ( \frac{\partial \mathcal{L}}{\partial (\partial_{\mu}\ub)} \right ) & = 0 &
    \implies \rho \ddot{u} & = \partial_z \left ( G_1 (\partial_{\bar{z}} u + \partial_{z} \bar{u}) \right )  -  \partial_{z} \left ( G_2 (\partial_z \ub - \partial_{\zb} u) \right ) + \mathcal{O}\left(B\right) +\mathcal{O}\left(M/N\right) 
    \label{SIeqn:complexEOM}
\end{align}
which relates change in momentum density ($\rho \ddot{u}$) to the internal force density coming from local stress, in complex notation. To accommodate external forces in our analysis, we can include the appropriate force density terms in the R.H.S., as is typical in Lagrangian mechanics.



\subsection{Low-frequency Conformal Modes}

The ideal conformal Lagrangian $\mathcal{L}_{\mathrm{conf}}$ [\cref{SIeqn:confL}] has a degenerate manifold of conformal modes found as zero-frequency stress-free solutions of the equation of motion, \cref{SIeqn:complexEOM}, in the ideal-dilational limit $B \to 0$ and $1/N \to 0$:
\begin{align}
   u(z,\zb,t) & = f(z)t + g(z) & \iff && \sigma = G_1 s_1 + iG_2 s_2 & = 0.
\end{align}
These modes combine a constant-speed deformation $f(z)t$ with a static displacement field $g(z)$, both complex-analytic to ensure a conformal (shear-free) mapping. These include the trivial uniform translation modes $f(z)=a + ib$ and uniform rotation mode $f(z)= iz$ expected in any isolated system. Fields such as $f(z)=z$ and those that are higher-order in $z$ describe new zero-frequency modes not found in conventional elasticity.

For the realistic case, with non-zero perturbations of order $B$ and $M/N$, these modes are lifted to small but finite frequencies with some non-conformal corrections to the conformal profiles. 
To zeroth order in the perturbative expansion, we find the zero-frequency conformal modes associated with exact conformal symmetry.
At first-order expansion in either perturbation parameter, say $\epsilon \in \{B/G_2, 1/N\}$, these conformal modes acquire a small non-conformal correction and a small frequency $\omega_m$:
\begin{align}
    u_m(z,\bar{z},t) & = \left ( u^{(0)}_m(z) + \epsilon u^{(1)}_m(z,\bar{z}) \right ) \cos(\omega_m t + \phi), \label{SIeqn:perturbexpansion}
\end{align}
where $m$ indexes in order of increasing frequency, and the squared mode frequency scales linearly with the perturbation parameters $\omega_m^2 \sim \mathcal{O}\left(\epsilon\right)$.

To find the purely conformal approximation to these modes, $u_m^{(0)}(z)$, we limit the dynamics to the space of complex-analytic displacement fields, i.e. zero-shear deformations $\partial_{\zb} u = 0$. Such fields can be taylor-expanded in $z$ as $u(z)=\sum_n c_n z^n$ with a suitable cutoff $n_c$ to ignore higher-order terms that generate strong enough dilational stress making shearing energetically viable. This reduces the dynamics of conformal modes to the space of polynomials of order $\leq n_c$:
\begin{align}
\begin{aligned}
    \mathrm{L}[u(z,t)] & = \int \mathcal{L} \, d^2 z =  \int \left ( \frac{\rho}{2} |\dot{u}(z)|^2 - 2B \mathrm{Re}[u'(z)]^2 - 2\frac{M}{N} L^2 |u''(z)|^2 \right )  \, d^2z,
\end{aligned}\label{SIeqn:conformallagrangian}
\end{align}
where $u(z,t) = \sum_{n=0}^{n_c} c_n(t) z^n$, and $c_n(t)$ are the time-dependent coefficients, to be found by extremising the associated action.

One may smartly change basis from $\{z^n\}_n$ to accommodate boundary conditions, and solve the reduced Lagrangian of \cref{SIeqn:conformallagrangian} in the new basis. For instance, to model our corner-clamped experimental setup, we incorporate the clamping constraints, $u(z_k)=0$ for each clamped point $z_k$, by choosing the basis:
\begin{align}
    \phi_n(z) & = z^n \prod_k (z - z_k) & \forall \,  0\leq n \leq n_c, \\
    u(z,t) & = \sum_{n=0}^{n_c} c_n(t) \phi_n(z).
\end{align}
The reduced Lagrangian can be represented as a quadratic form in $\mathbf{c}(t) = [c_0(t), c_1(t), c_2(t), \ldots]$ with constant coefficients
\begin{align}
    \mathrm{L}[\mathbf{c}, \dot{\mathbf{c}}]& = \sum_{n,m} \bigg (  \frac{\rho}{2} \langle \phi_n | \phi_m \rangle \dot{\bar{c}}_n \dot{c}_m   - \frac{B}{2} \left ( \langle \bar{\phi}'_n | \phi'_m \rangle c_n c_m  + 2 \langle \phi'_n | \phi'_m \rangle \bar{c}_n c_m   + \langle \phi'_n | \bar{\phi}'_m \rangle \bar{c}_n \bar{c}_m \right ) - 2\frac{M}{N} L^2 \bar{c}_n c_m \langle \phi''_n| \phi''_m \rangle \bigg ) \nonumber\\
    & =  \frac{\rho}{2} \dot{\mathbf{c}}^{\dagger} \mathrm{A}_0 \dot{\mathbf{c}} - \frac{B}{2} \left (  \mathbf{c}^{T} \mathrm{A}_1 \mathbf{c}  + 2 \mathbf{c}^{\dagger} \mathrm{A}_2 \mathbf{c}  + (\mathbf{c}^{T} \mathrm{A}_1 \mathbf{c})^* \right ) - 2\frac{M}{N} L^2  \mathbf{c}^{\dagger} \mathrm{A}_3 \mathbf{c}, 
\end{align}
where $A_i$ are matrices composed of overlap integrals of the chosen conformal basis and its derivatives:
\begin{align}
    (\mathrm{A}_0)_{nm} & = \langle \phi_n | \phi_m \rangle, & (\mathrm{A}_1)_{nm} & = \langle \bar{\phi}'_n | \phi'_m \rangle, & (\mathrm{A}_2)_{nm} & = \langle \phi'_n | \phi'_m \rangle, & (\mathrm{A}_3)_{nm} & = \langle \phi''_n| \phi''_m \rangle.
\end{align}
Solving this Lagrangian yields a linear system of second-order differential equations in the coefficient vector $\mathbf{c}(t)$,
\begin{align}
    \rho \mathrm{A}_0 \ddot{\mathbf{c}} & = 2B \left ( \mathrm{A}_1^{\dagger} \mathbf{c}^* + \mathrm{A}_2 \mathbf{c} \right ) + 4\frac{M}{N} L^2 \mathrm{A}_3 \mathbf{c}.
\end{align}
The eigenfunctions $\mathbf{c}(t)= \mathbf{c}_m e^{i\omega_m t} $ with eigenfrequencies $\omega_m$ are numerically calculated to find the leading-order conformal modes:
\begin{align}
    u_m^{(0)}(z,t) & = \cos(\omega_m t + \phi) \sum_n c_{mn} \phi_n(z) 
\end{align}
as reported in main-paper Fig.~4a,b with $n_c=20$. The mode frequencies roughly scale with mode number $m$ as
\begin{align}
    \omega_m^2 & \sim \frac{B}{\rho L^2} \mathcal{O}\left(m^2\right) +\frac{M}{N \rho L^2} \mathcal{O}\left(m^4\right).
\end{align}

\subsection{High-frequency non-conformal modes}

The finite frequency modes in the ideal limit are found by solving the wave equation \cref{SIeqn:complexEOM} for $B\to 0,\, 1/N \to 0$. These plane-wave modes are found to be of two orthogonal polarisations propagating with distinct speeds $c_1 = \sqrt{G_1/\rho}$ and $c_2 = \sqrt{G_2/\rho}$. Instead of the usual longitudinal and transverse acoustic polarisations, these correspond to having only one oscillating shear strain, $s_1$ or $s_2$, as described here using vector fields in real coordinates:
\begin{align}
    \du_1(x,y,t) & = a \begin{bmatrix}
        - k_x \\ k_y
    \end{bmatrix} \exp[i(\omega_1 t - \dk \cdot \dr)] & \omega_1(|\dk|) & = c_1 |\dk| & s_2 & = \partial_y u_x + \partial_x u_y = 0 \\
    \du_2(x,y,t) & = a\begin{bmatrix}
        k_y \\ k_x
    \end{bmatrix} \exp[i(\omega_2 t - \dk \cdot \dr)] & \omega_2(|\dk|) & = c_2 |\dk| & s_1 & = \partial_x u_x - \partial_y u_y = 0
\end{align}
where $\dk = (k_x, k_y)$ is the wave-vector.
For a rectangular system of size $L_x, L_y$ centred at $\dr=0$ with corners clamped to have zero movement, the allowed modes are standing waves indexed by non-zero integers $n,m$ with wavevector and frequency given by
\begin{align}
    \dk(n,m) & = \frac{n\pi}{L_x} \hat{x} + \frac{m\pi}{L_y} \hat{y}, & \omega_{i}(n,m) & = c_i \pi \sqrt{\frac{n^2}{L_x^2} + \frac{m^2}{L_y^2}}, \qquad i = 1,2.
\end{align}
The lowest-frequency non-conformal bulk modes have an $s_2$-polarisation with the smallest wave-vector, for $n=\pm 1$ and $m=\pm 1$, which sets the frequency ceiling for low-frequency conformal modes:
\begin{align}
    f_{\mathrm{ceil}} & = \frac{1}{2\pi} \omega_2 \left (1, 1 \right ) = \sqrt{\frac{G_2}{4\rho } \left ( \frac{1}{L_x^2} + \frac{1}{L_y^2} \right )},
\end{align}



\subsection{Conformal Symmetry \& Noether's Theorem}

A continuous symmetry is defined in terms of an infinitesimal transformation of the generalised fields and/or coordinates that leaves the Lagrangian density invariant. In other words, the dynamics are invariant under a specific continuous manifold of transformations. For an ideal-dilational material, the continuous conformal symmetry corresponds to conformal transformations that add an infinitesimal static conformal deformation (say $f(z)$) to an arbitrary trajectory: $u(z,\zb,t) \longmapsto u(z,\zb,t) + \epsilon f(z)$.
The conformal Lagrangian $\mathcal{L}_{\mathrm{conf}}$ [\cref{SIeqn:confL}] remains invariant under such transformations, since these don't change the local shear-strain $\partial_{\bar{z}} u \longmapsto \partial_{\bar{z}} u$, and the local velocities $\dot{u} \longmapsto \dot{u} $. Explicitly,
\begin{align}
    \delta_f \mathcal{L}_{\mathrm{conf}} &  =  \lim_{\epsilon \to 0} \frac{\mathcal{L}_{\mathrm{conf}}[u + \epsilon f] - \mathcal{L}_{\mathrm{conf}}[u]}{\epsilon}  = 0. \label{SIeqn:conformalinvariance}
\end{align}

For real dilational materials, modelled by the perturbed Lagrangian density \cref{SIeqn:perturbL}, conformal symmetry is broken by the perturbative terms that scale with $B$ and $1/N$:
\begin{align}
    \delta_f \mathcal{L} & = - 2 B \mathrm{Re}[\partial_z u] \mathrm{Re}[f'(z)] - \frac{M}{N} L^2 (\partial_z \mathrm{Re}[\partial_z u] \partial_{\zb}\mathrm{Re}[f'(z)] + \partial_{\zb} \mathrm{Re}[\partial_z u] \partial_{z}\mathrm{Re}[f'(z)])\label{SIeqn:approxinvariance} \\
    & = \mathcal{O}\left(Bf'\right) + \mathcal{O}\left(Lf'' M/N\right).
\end{align}
Thus, this symmetry is weakly broken for small $B$, large $N$ and slowly-varying generators $f(z)$.

\subsubsection{Local conservation law}

To apply Noether's Theorem and find conservation laws, we express $\delta_f \mathcal{L}$ in terms of its derivatives using chain rule
\begin{align}
    \delta_f \mathcal{L}
    & = \frac{\partial \mathcal{L}}{\partial (\partial_{\mu} u )} \partial_{\mu} f(z) + \frac{\partial \mathcal{L}}{\partial (\partial_{\mu}\ub)} \partial_{\mu} \overline{f(z)} \\
    & = \partial_{\mu} \left ( \frac{\partial \mathcal{L}}{\partial (\partial_{\mu} u )} f(z) \right ) - \partial_{\mu} \left ( \frac{\partial \mathcal{L} }{\partial (\partial_{\mu} u )} \right ) f(z) + \partial_{\mu} \left ( \frac{\partial \mathcal{L} }{\partial (\partial_{\mu} \ub )} \overline{f(z)} \right ) - \partial_{\mu} \left ( \frac{\partial \mathcal{L}}{\partial (\partial_{\mu} \ub )} \right ) \overline{f(z)} 
    \label{SIeqn:offshellcontinuity}
\end{align}
where, in the last step, we use the multiplication rule of differentiation to find the expressions of Euler-Lagrange equations \cref{SIeqn:eulerlagrangecomplex}.
The above is applicable to any trajectory $u(z,\zb,t)$ even if it is not physically possible, i.e. it may not minimise the action because it does not necessarily satisfy the Euler-Lagrange equations. To find conserved charges, we consider only physically possible dynamics, so called \emph{on-shell} trajectories, by enforcing the Euler-Lagrange equations, \cref{SIeqn:eulerlagrangecomplex}. This  eliminates the second and fourth terms in \cref{SIeqn:offshellcontinuity}:
\begin{align}
    \delta_f \mathcal{L} & = \partial_{\mu} \left ( \frac{\partial \mathcal{L}}{\partial (\partial_{\mu} u )} f(z) + \frac{\partial \mathcal{L}}{\partial (\partial_{\mu} \ub )} \overline{f(z)} \right ) \label{SIeqn:noetherlaw} \\
    \implies \mathcal{O}\left(Bf'\right) + \mathcal{O}\left(Lf'' M/N\right) & = \partial_{t} \left ( \frac{\partial T}{\partial \dot{u} } f(z) + \frac{\partial T}{\partial \dot{\ub} } \overline{f(z)} \right ) -  \partial_{i} \left ( \frac{\partial E}{\partial (\partial_{i} u) } f(z) + \frac{\partial E}{\partial (\partial_{i} \ub )} \overline{f(z)} \right )
\end{align}
where the perturbative terms are introduced using \cref{SIeqn:approxinvariance} and $i \in \{z,\zb\}$. Substituting the complex expressions for kinetic and potential energy densities, we find an approximate continuity relation for the momentum density $p_f$ along the corresponding conformal generator $f$:
\begin{align}
    \dot{p}_f + \partial_z j_f + \partial_{\bar{z}} \overline{j_f} & = \mathcal{O}\left(Bf'\right) + \mathcal{O}\left(Lf'' M/N\right) \label{SIeqn:contieqn} 
\end{align}
where $\rho_f(z,\zb,t)$ is defined as
\begin{align}
    p_f & := \partial_{t} \left ( \frac{\partial T}{\partial \dot{u}} f(z) + \frac{\partial T}{\partial \dot{\ub}} \overline{f(z)} \right ) = \mathrm{Re}[\rho \dot{u} \overline{f(z)}], \label{SIeqn:chargedensity}
\end{align}
and $j_f$ is the complex current density defined as
\begin{align}
    j_f & :=  - \frac{\partial E}{\partial (\partial_{i} u )} f(z) - \frac{\partial E}{\partial (\partial_{i} \ub )} \overline{f(z)}  = -(G_1 s_1 + iG_2 s_2) \overline{f(z)} - Bdf(z) + \mathcal{O}\left(f M/N\right)
\end{align}
where $s_1, s_2$ and $d$ are the linearised strains as expressed in \cref{SIeqn:complexstrains}. Note that $\sigma = G_1 s_1 + iG_2 s_2$ is the significant stress due to shear strains in complex notation.

\subsubsection{Global conservation law}
The total (global) Noether charge $Q_f$ corresponding to symmetry generator $f(z)$ is given by the area integral of the charge density $p_f$ over the spatial domain $D$, $Q_f := \int_D p_f \,d^2 z$, where $d^2 z = dxdy$ is an infinitesimal area element. Its rate of change is found using \cref{SIeqn:contieqn}
\begin{align}
    \dot{Q}_f + \int_D (\partial_z j_f + \partial_{\zb} \bar{j}_f) \, d^2 z & = \int_D \mathcal{O}\left(Bf'\right) + \mathcal{O}\left(Lf'' M/N\right) \, d^2 z. \label{SIeqn:changeinQf}
\end{align}
Using wedge products to denote the area element $d^2 z = dx \wedge dy = \frac{i}{2} dz \wedge d\zb$, we find that $d(j_f d\zb) = dj_f \wedge d\zb = \partial_z j_f dz \wedge d\zb = - 2i \partial_z j_f d^2 z $. Then, the complex Stokes' theorem is formulated as
\begin{align}
    \int_D \partial_z j_f d^2 z  & = \frac{i}{2}\int_{\partial D} j_f d\zb.
\end{align}
Adding this with its complex conjugate, yields the divergence integral seen in \cref{SIeqn:changeinQf} as a boundary flux integral:
\begin{align}
    \int_D (\partial_z j_f + \partial_{\zb} \bar{j}_f) d^2 z & = \frac{i}{2}\int_{\partial D} j_f d\zb - \frac{i}{2}\int_{\partial D} \bar{j}_f dz = -\int_{\partial D} \mathrm{Im} [j_f d\zb].
\end{align}
Finally, we have the approximate global conservation law given by
\begin{align}
    \dot{Q}_f  -\int_{\partial D}\mathrm{Im}[j_f d\zb] & = \int_D \mathcal{O}\left(Bf'\right) + \mathcal{O}\left(Lf'' M/N\right) \, d^2 z, \label{SIeqn:globallaw}
\end{align}
where the boundary integral of $j_f$ accounts for the net flow of conformal momentum $Q_f$ through $\partial D$. This assumes that the external forces only act at the boundary, which is true for our experiments. More generally, external forces may also act in the bulk and contribute additional sources or sinks of momentum change. The global conservation law states that, in the ideal limit, the total momentum $Q_f$ can only change due to external forces.



\section{Methods of Data Analysis}

\subsection{Conformal fitting of displacement fields}

To quantify the conformal nature of deformations seen in simulations and experiments, we fit a holomorphic function to the displacement fields using a least-square method. We consider deformation snapshots (or mode profiles) represented as a discrete complex displacement field $u[i], \, {i \in [N]}$ for $N$ blocks. The fitting function is expanded in a polynomial basis with suitably high order $n_c$ ($n_c=15$ in Fig.~4 and $n_c=10$ in Fig.~5),
\begin{align}
    f(z) & = \sum_{n=0}^{n_c} c_n \frac{z^n}{L^n}.
\end{align}
The normalisation by $L$ ensures that the basis functions $(z/L)^n$ remain $\mathcal{O}\left(1\right)$ across the system, preventing large variations in scale across polynomial order and improving the conditioning of the least-squares problem. With the known reference positions of the squares represented as complex coordinates $z_i,\,{i\in [N]}$, we minimize the squared error-in-fit:
\begin{align}
    \mathrm{Err}^2 & := \sum_{i\in [N]} \left | u[i] - f(z_i) \right |^2
\end{align}
with respect to the coefficient vector $\vec{c}=(c_n)_{n=0}^{n_c}$. This function is quadratic in the coefficients,
\begin{flalign}
    \mathrm{Err}^2 & = \vec{c}^{\dagger} A \vec{c} - \vec{c}^{\dagger} \vec{b} + \vec{b}^{\dagger} \vec{c} +  \sum_{i\in [N]} |u[i]|^2,
\end{flalign}
where $A_{nm} = \sum_{i\in [N]} \frac{\zb^n_i z^m_i}{L^{n+m}}$ are elements of matrix $A$, and $b_n = \sum_{i\in [N]} \frac{\zb^n_i u[i]}{L^{n}}$ are elements of vector $\vec{b}$. The error function is easily minimised when the following derivative is zero:
\begin{align}
    \frac{\partial \mathrm{Err}^2}{\partial \vec{c}^{\dagger}} & = A \vec{c} - \vec{b}= 0.
\end{align}
Thus, the coefficients for best fit are simply found as the solution of $A\vec{c}_{\mathrm{min}} = \vec{b}$. The corresponding conformal-fitting error is normalised by the norm of displacement field to define the measure of non-conformal deviation in the deformation $u$,
\begin{align}
    \Delta_{\mathrm{conf}}[u] & := \sqrt{\frac{\mathrm{Err}^2 \vert_{\vec{c}_{\mathrm{min}}}}{\sum_{i\in [N]} |u[i]|^2 }} = \sqrt{1 - \frac{\vec{b}^{\dagger} A^{-1} \vec{b}}{\sum_{i\in [N]} |u[i]|^2 }}.
\end{align}

\subsection{Verifying Conservation laws in Discrete RS Metamaterial}

To contend with the additional rotational degree of freedom in experimental and simulation data, we reformulate the continuum laws of conservation, \cref{SIeqn:globallaw}, in terms of discretised displacement-rotation fields. We define the discrete form of the conformal-symmetry generator $f(z)$ as a combination of the displacements $f(z_i^{\dn})$ for each square $i$ in unit cell $\dn$, together with the corresponding square rotations prescribed by the continuum approximation in \cref{SIeqn:goodbending,SIeqn:goodrotating}. We state this generator explicitly here in real coordinates:
\begin{flalign}
    && \vec{f} & :=(\mathbf{f}_i^{\dn},\phi_i^{\dn})_{i,\dn} && \\
    \text{where the displacement} && \mathbf{f}_i^{\dn} & = \left (\mathrm{Re}[f(z_i^{\dn})],\, \mathrm{Im}[f(z_i^{\dn})]\right ), \\
    \text{and the rotation} && \phi_i^{\dn} & = - \frac{k_la(a-l)\tan\theta_0}{8k_{\theta}+2k_l(a-l)^2\tan^2\theta_0} \mathrm{Re}[f'(z_i^{\dn})] -  \frac{(-1)^i}{2} \mathrm{Re}[f'(z_i^{\dn})].
\end{flalign}

Given an arbitrary trajectory $\vec{q}=  ( \du_{i}^{\dn}, \theta_{i}^{\dn})_{i,\dn}$, the infinitesimal symmetry operation is applied as $\vec{q} \longmapsto \vec{q} + \epsilon \vec{f}$, which affects the displacements and rotations as $\mathbf{u}_i^{\dn} \longmapsto \mathbf{u}_i^{\dn} + \epsilon\mathbf{f}_i^{\dn}$ and $\theta_i^{\dn} \longmapsto \theta_i^{\dn} + \epsilon\phi_i^{\dn}$ respectively.
The corresponding deviation in the discrete lattice Lagrangian, \cref{SIeqn:discretelagrangian}, is given by
\begin{align}
    \delta_f \mathrm{L} & = \delta_f T_{\mathrm{latt}} - \delta_f E_{\mathrm{latt}}.
\end{align}
Since, such a conformal generator is static in time ($\delta_f T_{\mathrm{latt}} = 0$), the Lagrangian remains invariant only if the generator does not cost deformation energy ($\delta_f E_{\mathrm{latt}} = 0$). From our continuum analysis in \Cref{sec:conformalmodel}, we know that these conformal displacements, with the associated counter-rotations, cost energy only at the order of the small effective bulk modulus and finite-size effects in the linear limit. Therefore,
\begin{align}
    \delta_f \mathrm{L} & = - \delta_f E_{\mathrm{latt}} = -\frac{\partial E_{\mathrm{latt}}}{\partial \vec{q}} \cdot \vec{f} \approx \int_{D} \mathcal{O}\left(f' B\right) + \mathcal{O}\left(L f'' M/N\right) \, d^2 z. \label{SIeqn:approxdiscreteinvariance}
\end{align}
Thus, we are able to reformulate the approximate conformal symmetry, shown in \cref{SIeqn:approxinvariance}, in the context of this discrete dilational lattice framework.

The corresponding Noether charge along $f$ is defined using the lattice kinetic energy, \cref{SIeqn:lattkineticquad}, similar to its continuum counterpart $p_f$ (see \cref{SIeqn:chargedensity}),
\begin{align}
    Q_f & := \vec{f} \cdot \frac{\delta T_{\mathrm{latt}}}{\delta \dot{\vec{q}}}  =  \vec{f} \cdot \mathcal{M} \dot{\vec{q}},\label{SIeqn:RS_momentum_defn}
\end{align}
where $\mathcal{M}$ is the inertia matrix. Its rate of change is found using the equation of motion, \cref{SIeqn:nonlin_EOM},
\begin{align}
    \dot{Q}_f = \vec{f} \cdot M \ddot{\vec{q}} & = - \frac{\partial E_{\mathrm{latt}}}{\partial \vec{q}} \cdot \vec{f} + \vec{F}_{\mathrm{ext}} \cdot \vec{f},
\end{align}
where we have included any external forces and torques applied on blocks as $\vec{F}_{\mathrm{ext}} = (\mathbf{F}_i^{\dn}, \tau_i^{\dn})_{i,\dn}$. The internal force component $- \frac{\partial E_{\mathrm{latt}}}{\partial \vec{q}} \cdot \vec{f}$ is equal to the symmetry-variation in Lagrangian: $\delta_f \mathrm{L} = - \delta_f E_{\mathrm{latt}} = - \frac{\partial E_{\mathrm{latt}}}{\partial \vec{q}} \cdot \vec{f}$. Using \cref{SIeqn:approxdiscreteinvariance}, we obtain the reformulated approximate conservation law for $Q_f$ in this discrete setting:
\begin{align}
    \dot{Q}_f - \vec{f} \cdot \vec{F}_{\mathrm{ext}} & \approx \int_{D} \mathcal{O}\left(f' B\right) + \mathcal{O}\left(L f'' M/N\right) \, d^2 z,
\end{align}
which becomes exact when bending stiffness is negligible ($B \to 0$) and lattice size large ($N \to \infty$), in absence of dissipation. In practice, internal dissipation in hinges introduces an additional symmetry-breaking term that is empirically smaller than the corrections above.

To verify approximate conservation laws, we calculate the deviation from exact conservation, $\dot{Q}_f - \vec{f} \cdot \vec{F}_{\mathrm{ext}}$, for experimental and simulation datasets, and normalise them according to the definition in eq. (13) of the main text to extract the inverse timescale of this deviation.


\bibliographystyle{plain}

\bibliography{refs}